\documentclass[final,journal,twoside]{IEEEtran}

\ifCLASSINFOpdf
  \usepackage[pdftex]{graphicx}
\else
  \usepackage[dvips]{graphicx}
\fi
\usepackage{mathrsfs}
\usepackage[cmex10]{amsmath}
\usepackage{amssymb,amsthm,amsfonts}
\usepackage{cite}
\usepackage{tikz}
\usepackage{threeparttable}

\newtheorem{theorem}{Theorem}
\theoremstyle{remark}
\newtheorem{remark}{Remark}
\newtheorem{example}{Example}

\newcommand{\const}[1]{{\mathcal{#1}}}
\newcommand{\Reals}{\mathbb{R}}
\newcommand{\Complex}{\mathbb{C}}
\newcommand{\normalr}[2]{\mathcal{N}_{\Reals}\!\left(#1,#2\right)}
\newcommand{\E}{\mathsf{E}}
\newcommand{\SNR}{\textnormal{SNR}}
\newcommand{\SNRs}{\textnormal{SNRs}}
\newcommand{\db}{\textnormal{dB}}
\newcommand{\ie}{\emph{i.e.}}
\newcommand{\eg}{\emph{e.g.}}
\newcommand{\der}{\mathrm{d}}
\newcommand{\inner}[2]{{\langle{#1},{#2}\rangle}}
\newcommand{\norm}[1]{\left\lVert#1\right\rVert}
\newcommand{\bigo}{\mathcal{O}}
\DeclareMathOperator{\diag}{diag}
\DeclareMathOperator{\offdiag}{offdiag}
\DeclareMathOperator{\sech}{sech}
\DeclareMathOperator{\sinc}{sinc}

\begin{document}

\title{Information Transmission using
the Nonlinear Fourier Transform, Part III:\\ Spectrum
Modulation\thanks{\noindent Submitted for publication on February 12,
  2013. The material in this paper was presented in part at the 2013 IEEE International Symposium on 
Information Theory. The authors are with the Edward S. Rogers Sr.\ Dept.\ of
Electrical and Computer Engineering, University of Toronto,
Toronto, ON M5S 3G4, Canada. Email:
\texttt{\{mansoor,frank\}@comm.utoronto.ca.}}}

\markboth{IEEE Transactions on Information Theory}{Yousefi and
Kschischang}

\author{Mansoor~I.~Yousefi and Frank~R.~Kschischang,~\IEEEmembership{Fellow, IEEE}}

\IEEEpubid{0000--0000/00\$00.00~\copyright~2013 IEEE}

\maketitle

\begin{abstract}
\ifCLASSOPTIONonecolumn\relax\else\boldmath\fi
Motivated by the looming ``capacity crunch'' in fiber-optic
networks, information transmission over such systems is
revisited.  Among numerous distortions, inter-channel
interference in multiuser wavelength-division multiplexing
(WDM) is identified as the seemingly intractable factor
limiting the achievable rate at high launch power.  However,
this distortion and similar ones arising from nonlinearity
are primarily due to the use of methods suited for linear
systems, namely WDM and linear pulse-train transmission, for
the nonlinear optical channel.  Exploiting the integrability
of the nonlinear Schr\"odinger (NLS) equation, a nonlinear
frequency-division multiplexing (NFDM) scheme is presented,
which directly modulates non-interacting signal
degrees-of-freedom under NLS propagation.  The main
distinction between this and previous methods is that NFDM
is able to cope with the nonlinearity, and thus, as the the
signal power or transmission distance is increased, the new
method does not suffer from the deterministic cross-talk
between signal components which has degraded the performance
of previous approaches.  In this paper, emphasis is placed
on modulation of the discrete component of the nonlinear
Fourier transform of the signal and some simple examples of
achievable spectral efficiencies are provided.
\end{abstract}

\ifCLASSOPTIONonecolumn\clearpage\else\relax\fi


\section{Introduction}
\label{sec:intro}
\IEEEPARstart{T}{his paper} is a continuation of Part~I
\cite{yousefi2012nft1} and Part~II \cite{yousefi2012nft2} on
data transmission using the nonlinear Fourier transform
(NFT). [Part~I] describes the mathematical tools underlying
this approach to communications. Numerical methods for
implementing the NFT at the receiver are discussed in
[Part~II].  The aims of this paper are to provide methods
for implementing the inverse NFT at the transmitter, to
discuss the influence of noise on the received spectra, and
to provide some example transmission schemes, which
illustrate some of the spectral efficiencies achievable by
this method.

The proposed nonlinear frequency-division multiplexing
(NFDM) scheme can be considered as a generalization of
orthogonal frequency-division multiplexing (OFDM) to
integrable nonlinear dispersive communication channels
\cite{yousefi2012nft1}. The advantages of NFDM stem from the
following:
\begin{enumerate}
\item NFDM removes deterministic inter-channel interference
(cross-talk) between users of a network sharing the same
fiber channel;
\item NFDM removes deterministic inter-symbol interference
(ISI) (intra-channel interactions) for each user;
\item spectral invariants as carriers of data are remarkably
stable and noise-robust features of the nonlinear
Schr\"odinger (NLS) flow;
\item with NFDM, information in each channel of interest can
be conveniently read anywhere in a network independently of
the optical path length(s).
\end{enumerate}

As described in [Part~I], the nonlinear Fourier transform of
a signal with respect to a Lax operator consists of discrete
and continuous spectral functions, in one-to-one
correspondence with the signal.  In this paper we focus
mainly on discrete spectrum modulation, which captures a
large class of input signals of interest.  For this class of
signals, the inverse NFT is a map from $2N$ complex
parameters (discrete spectral degrees-of-freedom) to an
$N$-soliton pulse in the time domain.  This special case
corresponds to an optical communication system employing
multi-soliton transmission and detection in the focusing
regime.

A physically important integrable channel is the optical
fiber channel. Despite substantial effort, fiber-optic
communications using fundamental solitons (\ie, 1-solitons)
has faced numerous challenges in the past decades.  This is
partly because the spectral efficiency of conventional
soliton systems is typically quite low ($\rho\sim 0.2$
bits/s/Hz), but also because on-off keyed solitons interact
with each other, and in the presence of noise the system
reach is limited by the Gordon-Haus effect
\cite{mollenauer2006sof}.  Although solutions have been
suggested to alleviate these limitations
\cite{mollenauer2006sof}, most current research is focused
on the use of spectrally-efficient pulse shapes, such as
sinc and raised-cosine pulses, with digital backpropagation
at the receiver \cite{essiambre2010clo}. Although these
approaches provide a substantial spectral efficiency at low
to moderate signal-to-noise ratios (\SNRs), their efficacy
saturates after a finite $\SNR\sim 20-30~\db$ where
$\rho\sim 5-9~\text{bits/s/Hz}$. This, as we shall see in
Section~\ref{sec:cap-limit}, is due to the incompatibility
of the wavelength-division multiplexing (WDM) with the flow
of the NLS equation, causing severe inter-channel
interference. 

\IEEEpubidadjcol

There is a vast body of literature on solitons in
mathematics, physics, and engineering; see, \eg,
\cite{ablowitz2006sai,faddeev2007hmi,mollenauer2006sof,hasegawa1995solitons,singer1996spc}
and references therein. Classical, path-averaged and
dispersion-managed fundamental 1-solitons are well-studied
in fiber optics \cite{mollenauer2006sof}.  The existence of
optical $N$-soliton pulses in optical fibers is also well
known \cite{mollenauer2006sof};  this previous work is
mostly confined to the pulse-propagation properties of
$N$-solitons, is usually limited to small $N$ (\eg,
$N=2,3$), or focuses on specific isolated $N$-solitons (\eg,
pulses of the form $A\sech(t)$). Signal processing problems
(\eg , detection and estimation) involving soliton signals
in the Toda lattice and other models have been considered by
Singer \cite{singer1996spc}. There is also a related work by
Hasegawa and Nyu, ``eigenvalue communication''
\cite{hasegawa1993ec}, which is reviewed and compared with
our approach in Section~\ref{sec:eig-com}, after NFDM is
explained.

While a fundamental soliton can be modulated, detected and
analyzed in the time domain, $N$-solitons are best
understood via their spectrum in the complex plane. In this
paper, these pulses are obtained by implementing a
simplified inverse NFT at the transmitter using the Darboux
transform and are demodulated at the receiver by recovering
their spectral content using the forward NFT. Since the
spectral parameters of a multi-soliton naturally do not
interact with one another (at least in the absence of
noise), there is potentially a great advantage in directly
modulating these non-interacting degrees-of-freedom. Sending
an $N$-soliton train for large $N$ and detecting it at the
receiver---a daunting task in the time domain due to the
interaction of the individual components---can be
efficiently accomplished, with the help of the NFT, in the
nonlinear frequency domain.
 
The paper is organized as follows. In Section
\ref{sec:cap-limit}, we revisit the wavelength-division
multiplexing method commonly used in in optical fiber
networks and identify inter-channel interference as the
capacity bottleneck in this method.  This section provides
further motivation for the NFT approach taken here.  In
Section~\ref{sec:disc-spec-fun}, we study algorithms for
implementing the inverse nonlinear Fourier transform at the
transmitter for signals having only a discrete spectrum.
Among several methods, the Darboux transform is found to
provide a suitable approach.  The first-order statistics of
the (discrete) eigenvalues and the continuous spectral
amplitudes in the presence of noise are calculated in
Section~\ref{sec:statistics}. In Section~\ref{sec:examples}
we calculate some spectral efficiencies achievable using
very simple NFT examples.  Finally, we provide some remarks
on the use of the NFT method in Section~\ref{sec:remarks}
and conclude the paper in Section~\ref{sec:conclusions}.

\section{Origin of Capacity Limits in WDM Optical Networks}
\label{sec:cap-limit} Recent studies on the capacity of WDM
optical fiber networks suggest that the information rates of
such networks is ultimately limited by the impacts of the
nonlinearity, namely inter-channel and intra-channel
nonlinear interactions \cite{mitra2000nli,
essiambre2010clo}. The distortions arising from these
interactions have deterministic and (signal-dependent)
stochastic components that grow with the input signal power,
diminishing the achievable rate\footnote{In this paper, the
term ``\emph{achievable rate}'' refers to a lower bound to
the capacity. It is obtained by optimizing mutual
information under some assumptions, \eg, considering a
sub-optimal transmitter and receiver or a method of
communication, a subset of all possible input distributions,
etc.} at high powers.  In these studies, for a class of ring
constellations, the achievable rates of the WDM method
increases with average input power power, $\const{P}$,
reaching a peak at a certain critical input power, and then
asymptotically vanishes as $\const{P}\rightarrow\infty$ (see
\eg, \cite{essiambre2010clo} and references therein).

In this section we briefly review WDM, the method commonly
used to multiplex many channels in practical optical fiber
systems.  We identify the origin of capacity limitations in
this model and explain that this method and similar ones,
which are borrowed from linear systems theory, are poorly
suited for efficient communication over nonlinear optical
fiber networks.  In particular, certain factors limiting the
achievable rates in the prior work are an artifact of these
methods (notably WDM) and may not be fundamental.  In
subsequent sections, we continue the development of the NFDM
approach that is able to overcome some of these limitations,
in a manner that is fundamentally compatible with the
structure of the nonlinear fiber-optic channel.

\subsection{System Model}

For convenience, we reproduce the system model given in
[Part~I].  We consider a standard single-mode fiber with
dispersion coefficient $\beta_2$, nonlinearity parameter
$\gamma$ and length $\const{L}$.  After appropriate
normalization (see, \eg, [Part~I, Section I, particularly
Eq. (3)]), the evolution of the slowly-varying part $q(t,z)$
of a narrowband signal as a function of retarded time $t$
and distance $z$ is well modeled by the stochastic nonlinear
Schr\"odinger equation
\begin{IEEEeqnarray}{rCl}
jq_z(t,z)=q_{tt}+2|q(t,z)|^2q(t,z) + n(t,z),
\label{eq:nnls}
\end{IEEEeqnarray}
where subscripts denote differentiation and $n(t,z)$ is a
bandlimited white Gaussian noise process, \ie , with
\[
\E \left\{n(t,z)n^*(t',z')\right\}
=\sigma^2\delta_{B}(t-t')\delta(z-z'),
\]
where $\delta_{B}(x)= 2B\mathrm{sinc}(2Bx)$, where $B$ is
the normalized noise bandwidth and where $\E$ denotes the
expected value.  It is assumed that the transmitter is
bandlimited to $B$ and power limited to $\const{P}$, \ie, 
\[
\E \frac{1}{\const{T}}\int_{0}^{\const{T}}|q(t,0)|^{2}dt=\const{P},
\] 
where $\const{T}\rightarrow\infty$ is the communication time. Note that the
power constraint in this paper is an equality constraint.

\begin{table}[b]
\caption{Fiber Parameters}
\label{tbl:fiberparam}
\centerline{\begin{tabular}{c|l|l}
$n_{\rm sp}$ & 1.1 & {\footnotesize excess spontaneous emission factor}\\
$h$ & $6.626 \times 10^{-34} {\rm J} \cdot {\rm s}$ & {\footnotesize Planck's constant} \\
$\nu$ & 193.55~{\rm THz} & {\footnotesize center frequency} \\
$\alpha$ & $0.046~{\rm km}^{-1}$ & {\footnotesize fiber loss (0.2~dB/km)} \\
$\gamma$ & $1.27~{\rm W}^{-1}{\rm km}^{-1}$ & {\footnotesize nonlinearity parameter} 
\end{tabular}}
\end{table}

In fiber-optic communication systems, noise can be introduced in a lumped
or distributed fashion.  The former case arises in systems
using erbium-doped fiber amplifiers (EDFAs) located at the
end of each fiber span \cite{mollenauer2006sof}.  We refer
to this type of noise as \emph{lumped noise}. If noise is
injected continuously throughout the fiber as a result of
distributed Raman amplification (DRA), as in
\eqref{eq:nnls}, one has \emph{distributed noise}
\cite{essiambre2010clo}. Here the fiber loss is assumed to
be perfectly compensated by the amplifier. In this paper, we
consider the DRA model, hence (\ref{eq:nnls}) explicitly
contains no loss term --- see also Remark~\ref{rem:loss}. In
this model, the noise spectral density is given by
$\sigma^2=\sigma_0^2\const{L}/(P_nT_n)$, where
$\sigma^2_0=n_{\rm sp}\alpha h\nu$, with parameters
given in Table~\ref{tbl:fiberparam}, and where
$P_n=2/(\gamma\const{L})$ and
$T_n=\sqrt{|\beta_2|\const{L}/2}$ are normalization
scale factors. The unnormalized signal and noise bandwidth
can be obtained by dividing the corresponding normalized
bandwidth by $T_n$.  A derivation of \eqref{eq:nnls} and a
discussion about sources of noise in fiber-optic channels
can be found in \cite{agrawal2001nfo,mollenauer2006sof}.

Mathematically, stochastic partial differential equations
(PDEs) such as \eqref{eq:nnls} are usually interpreted via
their equivalent integral representations. Integrating a
stochastic process with unbounded variation, such as white
noise, can be problematic. Consider, for instance, the
Riemann integral $\int_{z}^{z+\der z}g(z)\der B(z)\approx
g(l)(B(z+\der z)-B(z))$,  where $B(z)$ is the Wiener process
and $l\in[z,z+\der z]$. Since the integrand is not
approximately constant in $[z,z+\der z]$, the value of the
integral depends on $l$. The choice of $l$ leads to various
interpretations for a stochastic PDE, notably It\={o} and
Stratonovich representations, in which, respectively, $l=z$
and $l=z+\der z/2$. Fortunately, since in our application
noise is bandlimited in its temporal component, the
stochastic PDE \eqref{eq:nnls} is essentially a
finite-dimensional system and there is no difficulty in the
rigorous interpretation of \eqref{eq:nnls}.

\subsection{Achievable Rates of WDM Optical Fiber Networks}
\label{sec:wdm}
Fiber-optic communication systems often use
wavelength-division multiplexing to transmit information.
Similar to frequency-division multiplexing, information is
multiplexed in distinct wavelengths. This helps to separate
the signals of different users in a network, where they have
to share the same links between different nodes.

\begin{figure*}[t]
\centering
\includegraphics[width=\textwidth]{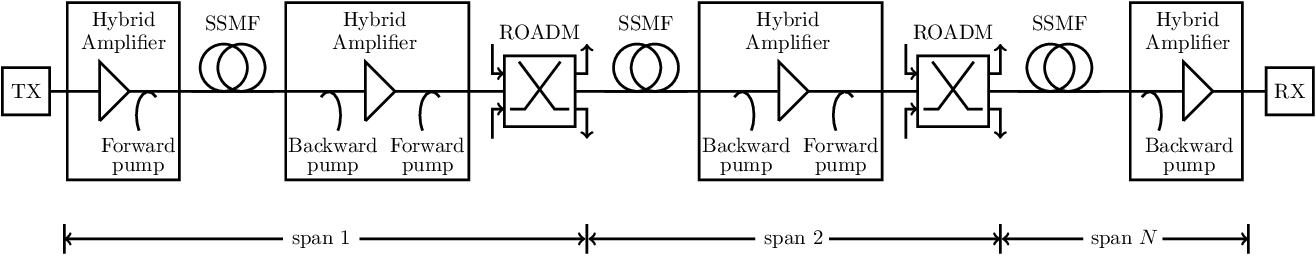}
\caption{Fiber-optic communication system (after
\cite{essiambre2010clo}).}
\label{fig:fiber-system}
\end{figure*}

Fig.~\ref{fig:fiber-system} shows the system model of a link
in an optical fiber network between a source and a
destination. There are $N$ fiber spans between multiple
users at the transmitter (TX) and multiple users at the
receiver (RX). The signal of some of these users is destined
to a receiver other than the RX shown in
Fig.\ref{fig:fiber-system}. As a result, at the end of each
span there is a reconfigurable optical add-drop multiplexer
(ROADM) that may drop the signal of some of the users or, if
there are unused frequency bands, add the signal of
potential external users.  We are interested in evaluating
the per degree-of-freedom capacity (bits/s/Hz) of the
optical fiber link from the transmitter to the receiver. 

In WDM, the following (baseband) signal is transmitted over the channel 
\begin{IEEEeqnarray}{rCl}
  q(t,0)=\sum\limits_{k=0}^{N-1}
\left(\sum\limits_{\ell=1}^{M}s_k^\ell\phi_\ell(t)\right)e^{j2\pi k W t},
\label{eq:wdm-signal}
\end{IEEEeqnarray}
where $k$ and $l$ are user and time indices,
$\{s_k^\ell\}_{l=1}^{M}$ are symbols transmitted by user
$k$, $W=B/N$ is the per-user bandwidth,
$\{\phi_\ell(t)\}_{l=1}^{\infty}$ is an orthonormal basis
for the space of finite energy signals with Fourier spectrum
in $[-W/2, W/2$], $N$ is the number of WDM users and $M$ is
number of symbols per user. To illustrate the essential
aspects, we temporarily simplify \eqref{eq:wdm-signal} by assuming that
$M=1$ and $\phi_1(t)=1$, $\forall t$, \ie, each user sends
a pure sinusoid, so as to work with Fourier series instead
of Fourier integrals. Thus each user operates at a single
frequency centered in a band of width $W$ and
\begin{equation}
 q(t,0)=\sum_{k=0}^{N-1} q_k(0) e^{j2\pi k W t} ,
 \label{eq:wdm-signal-z0}
\end{equation}
where $\{q_k(0)\}$ are the Fourier series coefficients at
$z=0$. 

As the periodic signal \eqref{eq:wdm-signal-z0} evolves in
the nonlinear optical fiber, new frequency components are
created and the signal may not remain periodic as in
\eqref{eq:wdm-signal-z0}. However, assuming a small $W$,
there are a large number of frequencies (users) at $z=0$ and
we can assume a Fourier series with variable coefficients
for $q(t,z)$ at  $z>0$
\begin{IEEEeqnarray}{rCl}
  q(t,z)=\sum\limits_{k=0}^{N-1} q_k(z)e^{j 2\pi k W t }.
\label{eq:fr}
\end{IEEEeqnarray}
Substituting \eqref{eq:fr} into
\eqref{eq:nnls}, we get the NLS equation in the discrete
frequency domain 
\begin{IEEEeqnarray}{rCl}
 j\frac{\partial q_k(z)}{\partial z}&=&\underbrace{-4\pi^2W^2
 k^2q_k(z)}_{\text{dispersion}}+
\underbrace{2|q_k(z)|^2q_k(z)}_{\textnormal{SPM}}\nonumber\\
&&+\underbrace{4q_k(z)\sum\limits_{\ell\neq
   k}|q_\ell(z)|^2}_{\textnormal{XPM}}\nonumber\\
&&+\underbrace{2\sum\limits_{\substack{\ell\neq m \\\ell\neq
     k}}q_\ell(z)q_m^*(z)q_{k+m-\ell}(z)}_{\textnormal{FWM}}+n_k(z),
\label{eq:freq-model}
\end{IEEEeqnarray}
in which $n_k$ are the noise coordinates in frequency and
where we have identified the dispersion, self-phase
modulation (SPM), cross-phase modulation (XPM) and four-wave
mixing (FWM) terms in the frequency domain\footnote{Some
authors define XPM differently.}. 

It is important to note that the optical WDM channel is a
nonlinear multiuser \emph{interference channel} with memory
\cite{el2011nit}. The inter-channel interference terms are
the XPM and FWM. There is no ISI in the assumed isolated
pulse transmission model \eqref{eq:freq-model} with one
degree-of-freedom per user.  However in a pulse-train
transmission model where $M>1$, replacing $\{q_k\}$ by
$\{s_k\}$ via the inverse transform shows that the other two
effects, the dispersion and SPM, cause inter-symbol
interference (intra-channel interaction).  Performance of a
WDM transmission system depends on how interference and ISI
are treated, and in particular the availability of the user
signals at the receiver. Several cases can be considered.

The received signal $q(t,\const{L})$ associated with \eqref{eq:wdm-signal} can be projected into
the space spanned by $\phi_l(t)\exp(j2\pi k W t)$,
$l=1,2,\ldots,M'$, $k=0,1,\ldots,N'-1$, for some $N'$ and $M'$, similar to
\eqref{eq:wdm-signal}. In this manner, the channel is
discretized as a map from a finite number of
degrees-of-freedom $s=\{s_k^\ell\}_{k,l=0,1}^{N-1,M}$ at the
channel input, to their corresponding values $\hat s=\{\hat
s_k^\ell\}_{k,l=0,1}^{N^\prime-1,M^\prime}$ at the channel
output. In general $N^\prime\neq N$ and $M^\prime\neq M$,
since signal bandwidth and duration at the transmitter and
receiver might be different.

If one has a finite number of degrees-of-freedom $s$  and $\hat s$,
has access to all of them and joint transmission and detection of $s$  and $\hat
s$ is practical, the channel is essentially a  single-user
vector channel $s\mapsto\hat s$, whose capacity is
non-decreasing with average input power~\cite{agrell2012mcf}
\[
\const{P} = \frac{1}{NM}\sum\limits_{k=0}^{N-1}\sum\limits_{l=1}^{M} \E
|s_k^\ell|^2.
\]
If joint transmission and detection is not possible, \eg, in
frequency $k$ or in time $\ell$, then one can either treat
interference as noise (as currently assumed in WDM
networks), or examine various strategies to manage
interference. By analogy with the linear $N$-user
interference channel \cite{kramer2006rrr,cadambe2008ia},
these strategies include signal-space orthogonalization
(\eg, as achieved by NFDM in the deterministic model) and,
if enough information about the channel is known,
interference cancellation (particularly in the strong
regime) and interference alignment.  If one of these
interference management strategies can be successfully
applied, then, again, the capacity of each user can be
non-decreasing with average input power.  If none of these
strategies is applicable so that interference is treated as
noise, or if additional constraints are present, the
achievable rate of the channel of interest in WDM can
saturate or decrease with average input power. Below, we clarify these cases in
more detail.

It is obvious that capacity is a non-decreasing function of
cost under an inequality constraint.  Below, we assume an
average cost defined by an equality constraint. This may not
be a suitable definition from a practical point of view, but
it is certainly of theoretical interest and it is also the
convention in optical fiber communication. 

\subsubsection{Single-user Memoryless Channels}

The capacity-cost function of a single-user vector
discrete-time memoryless channel with input
alphabet having a symbol with unbounded cost is a non-decreasing
function of an equality-constrained average cost
\cite{agrell2012mcf}.  The argument of \cite{agrell2012mcf}
goes as follows.  If a rate $R$ is achievable at cost
$\mathcal{P}$ by some input distribution $p(x)$, then, for a
small positive $\epsilon$, a rate of at least
$(1-\epsilon)R$ is achievable at cost
$\mathcal{P}'>\mathcal{P}$ using the distribution
$(1-\epsilon)p(x)+\epsilon \delta(x-x_1)$ where $x_1$ is a
symbol of large cost.  Intuitively, sending a symbol $x_1$
of large cost with small probability allows average cost to
grow with negligible impact on the achieved rate.  Since the
channel is memoryless, transmission of $x_1$ does not affect
the other symbols transmitted.  Essentially the transmitter
remains in a low-power state most of the time, which
effectively turns the equality constraint to an inequality
constraint.  See \cite{agrell2012mcf} for details, as well
as for a discussion about more general scenarios.

The monotonicity of $C(\const{P})$, of course, holds true
for any set of transition probabilities in a discrete-time
memoryless channel, including those obtained from nonlinear
channels.  Consider, for instance, a nonlinear memoryless
channel, \eg, the continuous-time zero-dispersion optical
fiber channel in the presence of a filter at the receiver.
When a signal propagates in this channel, its spectrum can
spread continuously. The amount of spectral broadening
depends on the pulse shape and, in particular, on the signal
intensity. Thus a signal with large cost may also require a
large transmission bandwidth and may be filtered out by the
receiver filter. However, a large-cost dummy signal does not
need to be decoded. Thus, as for any discrete-time
memoryless channel, the capacity (bits/symbol) is non-decreasing
with the average input power $\const{P}$.

Although $C(\mathcal{P})$ is monotonic, it may saturate,
\ie, approach a finite constant for large values of
$\mathcal{P}$.  In the zero-dispersion optical fiber
example, nonlinearity will cause a signal-dependent spectral
broadening, and large-energy signals may broaden beyond the
bandwidth of the receiver filter.  Thus the nonlinearity
could potentially cause the capacity $C(\mathcal{P})$ to
saturate;  a precise analysis would depend on the
definitions of bandwidth and time duration.

Of course a saturating $C(\const{P})$ is a serious
limitation to data communications. Firstly, from a practical
standpoint, a capacity that saturates is equivalent to one
that that peaks.  Secondly, in many channels one may not be
able to increase the average cost in the particular manner
described above.  For instance, it is not possible to send a
symbol with arbitrarily large cost in a channel in which
each symbol has a finite cost or in the presence of a peak-power
constraint. Thirdly, in some cases increasing the
average cost will limit the admissible input distributions
and decrease the capacity\footnote{As a simple example, a
binary-input channel with input costs $c_0<c_1$ can achieve
a capacity $C(c_a)$ with average cost $c_a$ in the range
$c_0\leq c_a\leq c_1$.   However, since $C(c_0)=C(c_1)=0$,
$C(c_a)$ is non-monotonic.  This situation can be observed
in computer simulations at average powers close to the peak
power.}.

\subsubsection{Single-user Channels with Memory}

The argument of \cite{agrell2012mcf} can be repeated for
channels with finite memory, \ie, when the influence of a
large-cost symbol vanishes in a finite time interval.
Whenever such a large-cost symbol is transmitted, the
receiver can simply wait for the channel to settle before
resuming normal operation.  As before, in the limit of small 
$\epsilon$, the loss in data rate is
negligible.  Of course, as before, saturation can occur; for
example, \cite{koch2009channelsheatup} gives an example of a
channel with memory where $C(\const{P})$ can saturate
even if optimal detection is performed.

In fiber-optic channels, the SPM term of each user is
available at the receiver for that user. Its deterministic
part, if needed, can be removed, \eg, by backpropagation and
its (signal-dependent) stochastic part can be handled by
coding and optimal detection over a long block of data (\eg,
maximum likelihood sequence detection). Deterministic or
stochastic nonlinear intra-channel effects do not cause the
capacity to vanish if the channel has finite memory (as in
the previous paragraph) and such joint detection is
performed.

Note that using a sub-optimal receiver in the fiber-optic
channel may cause the achievable rate to saturate with input
power. Consider, by way of analogy, a usual linear channel
$ y^m=T( s^m)+ n^m$, where $ s^m$, $ y^m$, $
n^m\in\Complex^m$ are, respectively, the input, output and
noise blocks and $T:\Complex^m\mapsto\Complex^m$ is an
invertible linear transformation. If $T$ is not a
multiplication (diagonal) operator, the channel is subject
to ISI.  Inverting the channel at the receiver completely
removes this ISI: $\hat{ s}_i=s_i+\hat n_i$, where $\hat{ s}
=T^{-1} y$ and $\hat{ n}=T^{-1} n$, giving an additive noise channel with colored, but signal-independent
noise.\footnote{Of course, naive channel inversion may
result in noise enhancement, which we ignore for the
purposes of this discussion.} A (suboptimal) receiver which
ignores noise correlation and performs isolated symbol
detection achieves a rate (lower bound to the capacity)
going to infinity with average input power.  In contrast,
now consider a nonlinear channel $  y^m=   F( s^m,  n^m)$,
where $F:\Complex^m\times\Complex^m\mapsto\Complex^m$ is a
nonlinear transformation, \ie, each output component $y_i$
is a nonlinear function of signal $s^m\in\Complex^m$ and
noise $n^m\in\Complex^m$. If $ F( s^m,0)$ is
invertible, channel inversion at the receiver, in general,
gives $\hat s_i=s_i+h_i( s^m, n^m)$, for some function
$h_i$.  As a result, in nonlinear systems, channel inversion
(\eg, by backpropagation) leaves a residual ``stochastic
ISI'' $h_i( s^m,  n^m)$ for each symbol.  This form of ISI
is absent when the noise is zero. In this case, a receiver
based on backpropagation and isolated symbol detection gives
rise to an ISI-limited communication system with  suboptimal
performance.  This occurs when assuming a memoryless model
for the fiber-optic channel. Treating stochastic ISI as
noise can lead to a bounded achievable rate. This is one
reason that $I(\const{P})$ has a peak in some of the prior
work.

The case of channels with infinite memory can be more
involved. Here sending a symbol with large cost may render
the rest of transmission useless. Thus care must be taken in
nonlinear channels in which the memory grows with signal.

\subsubsection{Multiuser Channels with Interference}
Inter-channel interference in the frequency domain is
mathematically the dual of intra-channel ISI in the time
domain. The difference is that 1) cooperation and joint
detection is generally not possible among users, and 2)
while the bandwidth is usually limited, transmission time
can be practically unlimited. In an optical fiber network,
many users have to share the same optical fiber link. Some
user signals join and leave the optical link at intermediate
points along the fiber, leaving behind a residual nonlinear
impact.  Thus we should assume that each user has access
only to the signal in its own frequency band, and the signal
of users $k^\prime \neq k$ is unknown to the user $k$.

In optical fiber networks, not much information is known
about interference signals. The location, or the number, of
ROADMs, and how many signals and with what properties have
joined or left the link may be unknown. These signals may
not co-originate, so cooperation and precoding among users
may not be possible.  In the absence of such information, it
is difficult to perform techniques such as interference
cancellation or alignment.  In this paper, we refer to such
system assumptions as the \emph{network scenario}. The WDM
simulations in the literature, and our analysis in this
section, assumes a network scenario, in which the nonlinear
interference is unavoidably treated as noise. 

In a nonlinear channel where, by definition, additivity is
not preserved under the action of the channel, multiplexing
user signals in a linear fashion, \eg, by adding them in
time (time-division multiplexing), in frequency (WDM or
traditional OFDM) or in space (multi-mode communications)
leads to inter-channel interference. In the case of WDM
where the available bandwidth is very limited, the
interference is significant and, if treated as noise in a
network scenario, ultimately limits the achievable rates of
such optical fiber systems. In
Section~\ref{sec:tdm-ofdm-mmf} a brief discussion of other
transmission techniques is given.

Among the two interference terms FWM and XPM, FWM is cubic
in signal amplitude and has a larger variance. It is obvious
that this interference grows rapidly when increasing the
common average power $\const{P}$ (or the number of users
$N$), ultimately overwhelming the signal and limiting the
achievable rate. The per-degree-of-freedom achievable rates
of the channel of interest in WDM method versus average power in a
network scenario, $I(\const{P})$, is noise-limited in the
low \SNR \ regime, following $\log(1+\SNR)$, and
interference-limited in the high \SNR\ regime, decreasing to
zero \cite{mitra2000nli, essiambre2010clo,goebelthesis,
mecozzi2012nonlinear}; see also Fig.~\ref{fig:5wdm-cap} and
Remark~\ref{rem:wdm-cap-clarification}.

To summarize the preceding discussion, the transmission
rates achievable over the nonlinear Schr\"odinger channel
depend on the method of transmission and detection, as well
as the assumptions on the model.  One can assume a
single-user or a multiuser channel,  with or without
memory, and with or without optical filtering.  It is thus
important, when comparing different results, to clarify
which modeling assumptions have been made.

\begin{figure*}[t]
\centering
\includegraphics[scale=1.0]{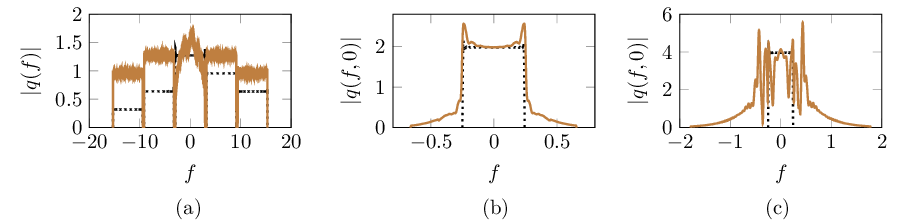} 
\caption{ (a) 5 WDM channels, with the channel of interest
at the center. The dotted and solid graphs represent, respectively,
the input and (noisy) output after backpropagation. Neighbor channels are dropped and added at
the end of the each span in the link, creating a residual interference
for the channel of interest. (b) Channel of interest at the input (dotted
rectangle) and at the output after backpropagation (solid
curve). The mismatch is due to the fact that the
backpropagation is performed only on the channel of interest
and the interference signals cannot be backpropagated. (c)
Inter-channel interference is increased with signal
intensity.}
\label{fig:interference}
\end{figure*}

\subsection{Inter-channel Interference as the Capacity Bottleneck in WDM Optical
Fiber Networks}

In the previous section we argued that, while intra-channel
interference can be handled by signal processing and coding,
inter-channel interference ultimately limits the achievable
rate of optical fiber networks.  The current practice in
fiber-optic communication is to send a linear sum of
signals in time (\eg, a pulse train) and in frequency (\eg,
WDM) in the form of \eqref{eq:wdm-signal}, the linear
orthogonality of which is corrupted by the nonlinear fiber
channel.  This corresponds to modulating
\emph{linear-algebraic modes} in the nonlinear channel (\eg,
sending sinc functions). Thus we identify conventional
linear multiplexing as a major culprit limiting the
achievable rate of current approaches in optical fiber
networks.  A modification of add-drop multiplexers is needed
so that the incoming signals are multiplexed in a nonlinear
fashion, exciting non-interacting signal degrees-of-freedom
under the NLS propagation. This corresponds to modulating
appropriate \emph{nonlinear modes} supported by the channel
(\eg, in the case of focusing regime, sending $N$-soliton
functions).

To illustrate the effect of the inter-channel interference
in the application of the WDM method to the nonlinear fiber
channel, we have simulated the transmission of 5 WDM
channels over $2000$~km of standard single-mode fiber with
parameters from Table~\ref{tbl:fiberparam}.  At the end of
each span, a ROADM filters the central channel of interest
(COI) at $40~\rm GHz$ bandwidth, and adds four independent
signals in neighboring bands, with symbols chosen uniformly
from a common constellation.  At the channel output, the COI
is filtered and backpropagated according to the inverse NLS
equation.  Fig.~\ref{fig:interference} compares the input
and output frequency-domain waveforms, after backpropagation
of the COI.  Fig.~\ref{fig:interference}(a) shows five
random instances of the multiplexed signals; note that,
because out-of-band signals are filtered and replaced at
various points along the fiber, the out-of-band signals at
the receiver are not related to the transmitted ones. Only
the COI is backpropagated.  Comparing
Fig.~\ref{fig:interference}(c) with
Fig.~\ref{fig:interference}(b), it can be seen that the
nonlinear inter-channel interference is stronger at higher
powers.

\begin{figure}[t]
\centering
\includegraphics{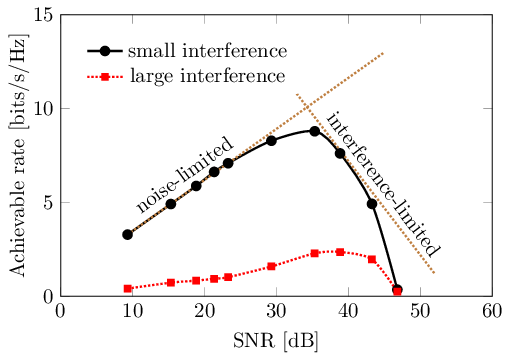} 
\caption{Achievable rates of the WDM method in a network scenario.}
\label{fig:5wdm-cap}
\end{figure}

The simulated achievable rates of WDM are shown in
Fig.~\ref{fig:5wdm-cap}. Here the distribution of the user
of interest is optimized and interference signals correspond
to independent symbols chosen uniformly from a common
multi-ring constellation. The two cases of large and small
inter-channel interference shown in the figure correspond to
large and small user peak powers, by scaling user
constellations. It is clear from both
Fig.~\ref{fig:interference} and Fig.~\ref{fig:5wdm-cap}
that, as the average transmitted power is increased, the
signal-to-noise ratio in the COI, and as a result the
information rate, vanishes to zero.  Note that this effect
can also be predicted by a simple \SNR\ analysis at the
receiver; see also \cite{ccbpf12}.

Using the mathematical and numerical tools described in
Parts I and II, this paper aims to show that it is possible
to exploit the integrability of the nonlinear Schr\"odinger
equation and induce a $k$-user interference channel on the
NLS equation so that both the \emph{deterministic}
inter-channel and inter-symbol interferences are
simultaneously zero for all users of a multiuser network.
Here by ``deterministic interference'' we mean interference
terms that are present even in the absence of noise.  This
lack of interference is a consequence of the integrability
of the cubic nonlinear Schr\"odinger equation in $1+1$
dimensions, and is generally not feasible for other types of
nonlinearity (even if the nonlinearity is weaker than
cubic!). This results in a deterministic
``orthogonalization'' for the nonlinear optical fiber
channel for any value of dispersion, nonlinearity, signal
power or transmission distance.

\begin{remark}
Note that, as a consequence of the data-processing
inequality, for an information-theoretic study it is not
necessary to perform deterministic signal processing such as
backpropagation. One is only concerned with transition
probabilities, which include effects such as rotations or
other deterministic transformations. Backpropagation just
aids the system engineer to simplify the task of the signal
recovery --- being an invertible operation, it
does not change the information content of the received
signal.
\end{remark}

\begin{remark}
\label{rem:wdm-cap-clarification}
An appropriate information-theoretic framework for WDM is to
describe the achievable rate region of the $N$-user
nonlinear interference channel with memory by a joint rate
$(R_1, \ldots, R_n)$. This is, however, difficult to
achieve. If one isolates a single channel, the corresponding
rate, $R_i$ , would depend on the distribution of the
signals of the other users (and not just their average
powers). In WDM, users can operate at low powers most of the
time, as prescribed in Section~\ref{sec:wdm} (a), and each
get a rate potentially saturating with power (even by
regarding interference as noise). However, in a network
scenario, interfering users may transmit data according to any
distribution --- including, in the extreme case, sending a
symbol with power $\const{P}$ all the time. The rates shown
in Fig.~\ref{fig:5wdm-cap}, vanishing at high powers, are
obtained when interfering users send data based on uniform
distributions, while the distribution of the user of interest
is optimized. The average power for each user was increased in 
the manner explained in the description of the simulation, and not
as in \cite{agrell2012mcf} (prescribed in Section~\ref{sec:wdm} (a)).
As noted earlier, this need not to be elaborated since the vanishing 
and saturating scenarios are essentially
equivalent. Some of the non-monotonic achievable rate graphs in the literature, similar
to Fig.~\ref{fig:5wdm-cap}, given appropriate assumptions, can be
interpreted as rates saturating with power at the location of the peak, by staying
in a low power regime most of the time, if needed.

\end{remark}

\section{The Discrete Spectral Function}
\label{sec:disc-spec-fun}
\subsection{Background}

Here we briefly recall the definition of the discrete
spectral function in the context of the nonlinear
Schr\"odinger equation. We first consider the deterministic
version of \eqref{eq:nnls}, where the noise is zero. Later,
we will treat noise as a perturbation of the noise-free
equation.

The nonlinear Fourier transform of a signal in
\eqref{eq:nnls} arises via the spectral analysis of the operator
\begin{IEEEeqnarray}{rCl}
L=j
\begin{pmatrix} 
\frac{\partial}{\partial t} & -q(t) \\
-q^*(t) & -\frac{\partial}{\partial t} 
\end{pmatrix}
=j\left(D\Sigma_3+Q\right),
\label{eq:L-op-part3}
\end{IEEEeqnarray}
where $D=\frac{\partial}{\partial t}$,
\begin{IEEEeqnarray*}{rCl}
Q=\begin{pmatrix}
0 & -q\\
-q^*& 0 
\end{pmatrix} \mbox{ and }
\Sigma_3=
\begin{pmatrix}
1 &  0 \\
0 & -1
\end{pmatrix}.
\end{IEEEeqnarray*}
Let $v(t,\lambda)$ be an eigenvector of $L$ with eigenvalue
$\lambda$.  Following [Part~I, Section~IV], the discrete spectral
function of the signal propagating according to
\eqref{eq:nnls} is obtained by solving the the
Zakharov-Shabat eigenproblem $Lv=\lambda v$, or equivalently
\begin{IEEEeqnarray}{rCl}
v_t=
\begin{pmatrix} 
-j\lambda & q(t) \\
-q^*(t) & j\lambda 
\end{pmatrix}v,\quad
v(t\rightarrow -\infty,\lambda)\rightarrow
\begin{pmatrix}1\\0\end{pmatrix}e^{-j\lambda t},
\IEEEeqnarraynumspace
\label{eq:dv-dt-part3}
\end{IEEEeqnarray}
where the initial condition was chosen based on the
assumption that the signal $q(t)$ vanishes as
$|t|\rightarrow \infty$. The system of ordinary differential
equations \eqref{eq:dv-dt-part3} is solved from $t=-\infty$
to $t=+\infty$ to obtain $v(+\infty,\lambda)$.  The
nonlinear Fourier coefficients $a(\lambda)$ and $b(\lambda)$
are then defined as
\begin{IEEEeqnarray*}{rCl}
a(\lambda)&=&\lim\limits_{t\rightarrow\infty}v_1(t,\lambda)e^{j\lambda t},\nonumber\\
b(\lambda)&=&\lim\limits_{t\rightarrow\infty}v_2(t,\lambda)e^{-j\lambda t}.
\end{IEEEeqnarray*}
Finally, the discrete spectral function is defined on the upper half
complex plane $\Complex^+=\{\lambda: \Im(\lambda)>0\}$:
\begin{IEEEeqnarray*}{rCl}
\tilde{q}(\lambda_j)=\frac{b(\lambda_j)}{ a_{\lambda}(\lambda_j)},\quad
j=1,\ldots,N,
\end{IEEEeqnarray*}
where subscript $\lambda$ denotes differentiation and $\lambda_j$ are the
isolated zeros of $a(\lambda)$ in
$\Complex^+$, \ie, solutions of $a(\lambda_j)=0$. The continuous
spectral function is defined on the real axis $\lambda\in\Reals$ as
$\hat{q}(\lambda)=b(\lambda)/a(\lambda)$.

\subsection{Modulating the Discrete Spectrum}

Let the nonlinear Fourier transform of the signal $q(t)$ be
represented by $q(t)\longleftrightarrow
(\hat{q}\left(\lambda), \tilde{q}(\lambda_j)\right)$. When
the continuous spectrum $\hat{q}(\lambda)$ is set to zero,
the nonlinear Fourier transform consists only of discrete
spectral functions $\tilde{q}(\lambda_j)$, \ie, $N$ complex
numbers $\lambda_1,\ldots, \lambda_N$ in $\Complex^+$
together with the corresponding $N$ complex spectral
amplitudes $\tilde{q}(\lambda_1), \ldots,
\tilde{q}(\lambda_N)$. In this case, the inverse nonlinear
Fourier transform can be worked out in closed-form, giving
rise to $N$-soliton pulses \cite{hirota2004direct}. The
simplified expressions, however, quickly get complicated
when $N>2$, and tend to be limited to low-order solitons. 

One can, however, create and modulate these multi-solitons
numerically.  In this section we study various schemes for
the implementation of the inverse NFT at the transmitter
when $\hat{q}=0$.  

\subsubsection{Discrete Spectrum Modulation by Solving
the Riemann-Hilbert System}

The inverse nonlinear Fourier transform can be obtained by
solving a Riemann-Hilbert system of integro-algebraic
equations or, alternatively, by solving the
Gelfand-Levitan-Marchenko integral equations --- see \eg,
[Part~I, Section~VII. A-B, particularly Eqs. (30a)--(30d)].
Great simplifications occur when $\hat{q}(\lambda)$ is zero.
For instance, in this case the integral terms in the
Riemann-Hilbert system vanish and the integro-algebraic
system of equations is reduced to an algebraic linear
system, whose solutions are $N$-soliton signals.

Let $V(t,\lambda_j)$ and $\tilde{V}(t,\lambda^*_j)$ denote
the scaled eigenvectors associated with $\lambda_j$ and
$\lambda_j^*$ defined by their boundary conditions at
$+\infty$ (they are denoted by $V^1$ and $\tilde{V}^1$ in
[Part~I]). Setting the continuous spectral function
$\hat{q}(\lambda)$ to zero in the Riemann-Hilbert system of
[Part~I, Eqs. (30a)--(30d)], we obtain an algebraic system of
equations
\begin{IEEEeqnarray}{rCl}
\tilde{V}(t,\lambda_m^*)&=&\begin{pmatrix}1\\0\end{pmatrix}+\sum\limits_{i=1}^{N}\frac{\tilde{q}(\lambda_i)e^{2j\lambda_i
    t}V(t,\lambda_i)}{\lambda_m^*-\lambda_i},\nonumber\\
\nonumber \\
V(t,\lambda_m)&=&\begin{pmatrix}0\\1\end{pmatrix}-\sum\limits_{i=1}^{N}\frac{\tilde{q}^*(\lambda_i)e^{-2j\lambda_i^*
    t}\tilde{V}(t,\lambda_i^*)}{\lambda_m-\lambda_i^*}.
\IEEEeqnarraynumspace
\label{eq:rh-integ-eq1}
\end{IEEEeqnarray}
Let $K$ be an $N\times N$ matrix with entries
\[
[K]_{ij}=\frac{\tilde{q}_ie^{2j\lambda_i t}}{\lambda_j^*-\lambda_i},\quad 1\leq i,j\leq N.
\]
Let $e_{N\times 1}$ be the all one column vector, $e_i=1$, $i=1,\cdots,N$,
and define variables
\begin{IEEEeqnarray*}{rCl}
 U_{2\times N}&=&\begin{pmatrix}V(t,\lambda_1) & V(t,\lambda_2) & \cdots&
   V(t,\lambda_N) \end{pmatrix}, \\
\tilde{U}_{2\times N}&=&\begin{pmatrix}\tilde{V}(t,\lambda_1^*) & 
\tilde{V}(t,\lambda_2^*) & \cdots & \tilde{V}(t,\lambda_N^*)\end{pmatrix},
\end{IEEEeqnarray*}
\begin{IEEEeqnarray*}{rCl}
(J_1)_{2\times N}&=&\begin{pmatrix} e^T \\
0
\end{pmatrix},
\quad
(J_2)_{2\times N}=\begin{pmatrix}  0 \\
e^T
\end{pmatrix},\\
\\
J_{2\times
  N}&=&J_2-J_1K^*=\begin{pmatrix}-e^{T}K^*\\e^T\end{pmatrix},\\
\tilde{J}_{2\times
  N}&=&J_1+J_2K=\begin{pmatrix}e^{T}\\e^TK\end{pmatrix},\\
\\
F_{N \times 1}&=&\begin{pmatrix}
\tilde{q}_1 e^{2j\lambda_1 t} & \tilde{q}_2 e^{2j\lambda_2 t} & \cdots & \tilde{q}_N e^{2j\lambda_N t} 
\end{pmatrix}^{T}.
\end{IEEEeqnarray*}  
Using these variables, the algebraic equations
\eqref{eq:rh-integ-eq1} are simplified to
\begin{IEEEeqnarray*}{rCl}
\tilde{U}=J_1+U K,  \quad \quad U= J_2-\tilde{U}K^*.
\end{IEEEeqnarray*}
Note that $K^*$ is the complex conjugate of
$K$ (not the conjugate transpose).  The solution of the above system is
\begin{IEEEeqnarray*}{rCl}
U&=&\left(J_2-J_1K^*\right)\left(I_N+KK^*\right)^{-1}=
J\left(I_N+KK^*\right)^{-1},
\\
\tilde{U}&=&\left(J_1+J_2K\right)\left(I_N+KK^*\right)^{-1}=
\tilde{J}\left(I_N+KK^*\right)^{-1}.
\end{IEEEeqnarray*}
From [Part I, Section. VII. B, Eq. 32], the $N$-soliton formula is given by
\begin{IEEEeqnarray}{rCl}
q(t)=-2j e^T
\left(I_N+K^*K\right)^{-1}F^*.
\label{eq:nsoliton-rh}
\end{IEEEeqnarray}
The right hand side is a complex scalar and has to be
evaluated for every $t$ to determine the value of $q(t)$
everywhere. 

\begin{example}
\label{ex:single-soliton}
It is useful to see that the (scaled) eigenvector for a single-soliton 
with spectrum
$\tilde{q}(\frac{\alpha+j\omega}{2},z)=\tilde{q}_0e^{2\alpha\omega
z}e^{-j(\alpha^2-\omega^2)z}$ is
\begin{IEEEeqnarray*}{rCl}
v(t,\lambda; z)=\frac{1}{2}\sech [\omega(t-t_0)]
\begin{pmatrix}
e^{-j\Phi}
\\
e^{\omega(t-t_0)}
\end{pmatrix},
\end{IEEEeqnarray*}
where $\Phi=\alpha
t+(\alpha^2-\omega^2)z-\angle\tilde{q}_0-\frac{\pi}{2}$ and
$t_0=\frac{1}{\omega}\log\left|\frac{\tilde{q}_0}{\omega}\right|-2\alpha
z$. The celebrated equation for the single-soliton obtained
from \eqref{eq:nsoliton-rh} is
\begin{IEEEeqnarray}{rCl}
q(t)=-j\omega e^{-j\alpha t}e^{-j\phi_0}\sech
\left(\omega (t- t_0)\right).
\label{eq:single-soliton}
\end{IEEEeqnarray}
From the phase-symmetry of the NLS equation, the factor $-j$
in \eqref{eq:single-soliton} can be dropped. The real and
imaginary part of the eigenvalue are the frequency and
amplitude of the soliton. Note that the discrete spectral
amplitude $\tilde{q}(\lambda)$ is responsible for the phase
and time-center of the soliton.
\end{example} 

Unfortunately the Riemann-Hilbert system is found to be
occasionally ill-conditioned for large $N$. The
$N^\text{th}$ row of the $K$ matrix is proportional to
$\exp(2j\lambda_N t)$. Thus this row gets a large scale
factor as $\Im(\lambda)$ is increased ($t<0$) which then
makes $I_N+K^* K$ ill-conditioned at large negative times.
As a result, the Riemann-Hilbert system, at least in the
current form, is not the best method for numerical
generation of $N$-solitons.

\subsubsection{Discrete Spectrum Modulation via the Hirota
Bilinearization Scheme}

It is also possible to generate multi-solitons without
solving a Riemann-Hilbert system or directly using the NFT.
A method which is particularly analytically insightful is
the Hirota direct method \cite{hirota2004direct}. It
prescribes, in some sense, a \emph{nonlinear superposition}
for integrable equations.

The Hirota method for an integrable equation works by
introducing a transformation of the dependent variable $q$
to convert the original nonlinear equation to one or more
\emph{homogeneous bilinear} PDEs. For integrable equations,
the nonlinearity usually is canceled or separated out. The
resulting bilinear equations have solutions that can be
expressed as sums of exponentials.  Computationally,
bilinear equations are solved perturbatively by expanding
the unknowns in terms of the powers of a small parameter
$\epsilon$. For integrable equations, this series truncates,
rendering approximate solutions of various orders to be
indeed exact.  The bilinear transformation has been found
for many integrable equations \cite{hirota2004direct},
taking on similar forms that usually involve the derivatives
of the logarithm of the transformed variable.

Let us substitute $q(t,z)=G(t,z)/F(t,z)$, where, without
loss of generality, we may assume that $F(t,z)$ is
real-valued.  To keep track of the effect of nonlinearity,
let us restore the nonlinearity parameter $\gamma$ in the
NLS equation.  Plugging $q=G/F$ into the NLS equation
\begin{IEEEeqnarray*}{rCl}
  jq_z=q_{tt}+2\gamma|q|^2q,
\end{IEEEeqnarray*}
we get 
\begin{IEEEeqnarray}{rCl}
j\left(G_zF-FG_z\right)&=&FG_{tt}-2F_tG_t-GF_{tt}\nonumber\\ 
&&+\: 2\frac{F_t^2+\gamma|G|^2}{F}G\nonumber\\
&=&FG_{tt}-2F_tG_t+GF_{tt}\nonumber\\ 
&&+\: 2\frac{F_t^2+\gamma|G|^2-FF_{tt}}{F}G,
\label{eq:trilinear}
\end{IEEEeqnarray}
where we have added and subtracted $2GF_{tt}$. Equation
\eqref{eq:trilinear} is trilinear in $F$ and $G$.  It can be
made bilinear by setting
\begin{IEEEeqnarray}{rCl}
j(G_zF-GF_z) &=& FG_{tt}-2F_tG_t+GF_{tt},\IEEEyesnumber
\IEEEyessubnumber \label{eq:bilinears-a}\\
F_t^2+\gamma|G|^2-FF_{tt}&=&0.\IEEEyessubnumber
\label{eq:bilinears-b}
\end{IEEEeqnarray}
This is a special solution for \eqref{eq:trilinear} which,
as we shall see, corresponds to $N$-soliton solutions. It is
very convenient (though not necessary) to organize
\eqref{eq:bilinears-a}--\eqref{eq:bilinears-b} using the Hirota $D$ operator
\begin{IEEEeqnarray*}{rCl}
  D_t^n(a(t),b(t))=\left(\frac{\partial }{\partial t}-\frac{
    \partial}{\partial t^\prime}\right)^n
    \left.a(t)b(t^\prime)\right|_{t^\prime=t},
\end{IEEEeqnarray*}
resulting in
\begin{IEEEeqnarray}{rCl}
(jD_z +D_t^2)FG&=&0,\IEEEyesnumber\IEEEyessubnumber\label{eq:FG1}\\
D_t^2 FF &=& 2\gamma |G|^2.\IEEEyessubnumber\label{eq:FG2}
\end{IEEEeqnarray}

Note that the $D$-operator acts on a pair of functions to
produce another function.  Note further that \eqref{eq:FG1}
does not depend on the nonlinearity parameter $\gamma$. That
is to say, the nonlinearity has been separated from equation
\eqref{eq:FG1}.  For some other integrable equations (\eg,
the Korteweg-de Vries equation) for which one gets only one
bilinear PDE, the nonlinearity parameter is in fact
canceled completely.

\begin{figure*}[t]
\centerline{\includegraphics{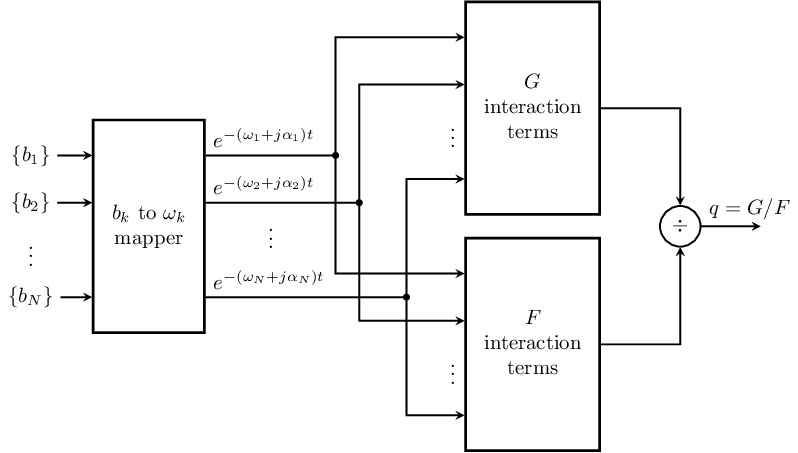}}
\caption{Hirota modulator for creating $N$-solitons.}
\label{fig:hirota}
 \end{figure*}

Bilinear Hirota equations \eqref{eq:FG1}--\eqref{eq:FG2} have solutions
in the form of a sum of exponentials. As shown in
Appendix~\ref{app:hirota}, $F$ and $G$ are obtained as \cite{hirota2004direct,osborne2009now}:
\begin{IEEEeqnarray*}{rCl}
  F(t,z) &=&
  \sum\limits_{b=\{0,1\}^{2N}}\delta_1(b)\exp\left(b^TX+b^TRb\right),\\
  G(t,z) &=&
  \sum\limits_{b=\{0,1\}^{2N}}\delta_2(b)\exp\left(b^TX+b^TRb \right),
\end{IEEEeqnarray*}
where $b=[b_i]_{i=1}^{2N}$ is a binary column vector, $b_i=\{0,1\}$,
$X=[X_i]_{i=1}^{2N}$, $X_i=\zeta_i t-k_iz+\phi_i$, $\zeta_{i+N}=\zeta_i^*$, $k_{i+N}=k_i^*$, $\phi_{i+N}=\phi_i^*$, $X_{i+N}=X_i^*$, 
 $k_i=j\zeta_i^2$
is the dispersion relation, $R_{2N\times 2N}$ is the Riemann matrix, 
\begin{IEEEeqnarray*}{rCl}
R_{ij}=
\begin{cases}
0, & i\geq j,\\
2\log\left(\zeta_i-\zeta_j\right)-\log\gamma, & \!\!\!\!\!\!\!\!(N+\frac{1}{2}-i)(N+\frac{1}{2}-j)>0, \\
-2\log\left(\zeta_i+\zeta_j\right)+\log\gamma, & i\leq N \mbox{ and } j\geq N+1 ,
\end{cases}
\IEEEeqnarraynumspace
\end{IEEEeqnarray*}
and
\begin{IEEEeqnarray*}{rCl}
  \delta_1(b)&=&
\begin{cases}
1, & \sum\limits_{i=1}^{N} b_i=\sum\limits_{i=N+1}^{2N}b_{i},\\
0, & \textnormal{otherwise},
\end{cases}
\\
  \delta_2(b)&=&
\begin{cases}
1, & \sum\limits_{i=1}^{N}b_{i}=1+\sum\limits_{i=N+1}^{2N} b_{i},\\
0, & \textnormal{otherwise}.
\end{cases}
\end{IEEEeqnarray*}
Matrix $R$ is upper triangular (with zeros on diagonal) and
such that when it is partitioned into four $N\times N$
blocks, the $11$, $12$ and $22$ blocks capture,
respectively,  the interaction between $X_i$ and $X_j$,
$X_i$ and $X_j^*$, and $X_i^*$ and $X_j^*$ variables. The
entries in the $11$, $12$ and $22$ blocks are, respectively,
given by $2\log(\zeta_i-\zeta_j)-\log\gamma$,
$-2\log(\zeta_i+\zeta_{N-j}^*)+\log\gamma$ and
$2\log(\zeta_{N-i}^*-\zeta_{N-j}^*)-\log\gamma$.  Eigenvalues
$\lambda_i$ are related to $\zeta_i$ via
$\zeta_i=-2j\lambda_i$, whereas the Hirota spectral
amplitudes $\phi_i$ are generally different from those of
other methods.

Functions $F$ and $G$ are in the form of the sum of all possible
exponentials such that in $F$ the number of non-conjugate
and conjugate variables $X_i$ and $X_j^*$ is the same while
in $G$ the former is one more than the latter. For each
exponential term, terms $R_{ij}$ corresponding to the
interaction between all possible pairs $X_{i}$ and
$X_{j}$, $i\neq j$, in the exponent are added; see
Table~\ref{tab:FG}.

Note that functions $F$ and $G$ both contribute to the
signal amplitude, whereas $F$, being real-valued, does not
contribute to the signal phase. Using the identity
$\partial_{tt}\log F=\left(F_{tt}F-F_t^2\right)/F^2$,
\eqref{eq:FG2} is reduced to
$|q(t,z)|^2=\gamma^{-1}\partial_{tt}\log F$. Therefore it is
also possible to derive the amplitude of $q$ solely in terms
of a function of $F$. This is because $F$ and $G$ are not
independent. 

Two important observations follow from the Hirota method.
Firstly, multi-soliton solutions of the NLS equation in the
$F$ and $G$ domain ($q=G/F$) are the summation of
exponentially decaying/growing functions $e^{\pm\omega t}
e^{j\alpha t-kz}$, each located at a frequency $\alpha$.
That is to say, while plane waves $e^{j \alpha t- k z}$ are
the natural Fourier basis functions that solve linear PDEs, for
integrable systems, exponentially decaying/growing functions
are suitable. The addition of the decaying/growing factor
$e^{\pm\omega t}$ is the point at which the nonlinear
Fourier transform diverges from the linear Fourier transform
\cite{osborne2009now}. Secondly, for each individual soliton
term, the Hirota method adds two-way interaction terms,
three-way interaction terms, etc., until all the
interactions are accounted for. In this way, the
interference between individual components is removed, as
shown schematically in Fig.~\ref{fig:hirota}.  Table
\ref{tab:FG} shows these interaction terms for $N=1,2,3$.

\begin{table*}[t]
\centering
\begin{threeparttable}[flushleft]
\caption{The Structure of the Interaction Terms in $F$ and $G$\tnote{1}.}
\begin{tabular}{r|c}
 & $F$\\ 
\hline
\rule{0pt}{4ex}$N=1$ & $1+e^{X_1+X_1^*+R_{13}}$ \\
\hline
\rule{0pt}{4ex}$2$ &
$1+e^{X_1+X_1^*+R_{13}}+e^{X_2+X_2^*+R_{24}}+e^{X_1+X_2^*+R_{14}}+e^{X_2+X_1^*+R_{23}}+e^{X_1+X_2+X_1^*+X_2^*+R_{12}+R_{13}+R_{14}+R_{23}+R_{24}+R_{34}}$ 
\\
\hline
\rule{0pt}{4ex}$3$ &
$1+e^{X_1+X_1^*}+e^{X_2+X_2^*}+e^{X_3+X_3^*}+\left(e^{X_1+X_2^*}+e^{X_2+X_1^*}+e^{X_1+X_3^*}+e^{X_3+X_1^*}+e^{X_2+X_3^*}+e^{X_3+X_2^*}\right)+$
\\
&
$\left(e^{X_1+X_2+X_1^*+X_2^*}+e^{X_1+X_2+X_1^*+X_3^*}+e^{X_1+X_2+X_2^*+X_3^*}+e^{X_1+X_3+X_1^*+X_2^*}+e^{X_1+X_3+X_1^*+X_3^*}+e^{X_1+X_3+X_2^*+X_3^*}\right.$
\\
&
$\left.+e^{X_2+X_3+X_1^*+X_3^*}+e^{X_2+X_3+X_1^*+X_2^*}+e^{X_2+X_3+X_2^*+X_3^*}\right)+e^{X_1+X_2+X_3+X_1^*+X_2^*+X_3^*}$\\\hline\hline
\rule{0pt}{4ex} & $G$\\ \hline
\rule{0pt}{4ex}$N=1$ & $e^{X_1}$ \\
\hline
\rule{0pt}{4ex}$2$ & $e^{X_1}+e^{X_2}+e^{X_1+X_2+X_1^*+R_{12}+R_{13}+R_{23}}+e^{X_1+X_2+X_2^*+R_{12}+R_{14}+R_{24}}$ 
\\
\hline
\rule{0pt}{4ex}$3$ &
$e^{X_1}+e^{X_2}+e^{X_3}+\left(e^{X_1+X_2+X_1^*}+e^{X_1+X_2+X_2^*}+e^{X_1+X_2+X_3^*}+e^{X_1+X_3+X_1^*}+e^{X_1+X_3+X_2^*}+e^{X_1+X_3+X_3^*}\right.$
\\
&
$\left.e^{X_2+X_3+X_1^*}+e^{X_2+X_3+X_2^*}+e^{X_2+X_3+X_3^*}\right)+\left(e^{X_1+X_2+X_3+X_1^*+X_2^*}+e^{X_1+X_2+X_3+X_1^*+X_3^*}+e^{X_1+X_2+X_3+X_2^*+X_3^*}\right)$
\end{tabular}
\label{tab:FG}
\tablenotes{
\small
\item [1]
For $N=3$, terms $R_{ij}$ in the exponent are not shown, due to space limitation.
}
\end{threeparttable}
\end{table*}

While the Hirota method reveals important facts about signal
degrees-of-freedom in the NLS equation, it may not be the
best method to compute multi-solitons numerically. There are
$\binom{2N}{N}\sim 2^{2N}$ and $\binom{2N}{N+1}\sim 2^{2N}$
terms in $F$ and $G$ respectively, and unless one truncates
the interaction terms at some step, the complexity quickly
grows, making it hard to compute $N$-solitons for $N > 10$.

\subsubsection{Recursive Discrete Spectrum Modulation Using
Darboux Transformation}

Multi-soliton solutions of the NLS equation can be
constructed recursively using the Darboux transformation. The Darboux transformation, originally introduced in
the context of the Sturm-Liouville differential equations
and later used in nonlinear integrable systems, provides the
possibility to construct a solution of an integrable
equation from another solution \cite{matveev1991darboux}.
For instance, one can start from the trivial solution $q=0$
of the NLS equation, and recursively obtain all higher-order
$N$-soliton solutions. This approach is particularly well
suited for numerical implementation.

Let $x(t,\lambda; q)$ denote a solution of the system
\begin{IEEEeqnarray}{rCl}
x_t&=&P(\zeta,q)x,\nonumber\\
x_z&=&M(\zeta,q)x,
\label{eq:dwdt-dwdz}
\end{IEEEeqnarray}
for the signal $q$ and complex number $\zeta=\lambda$ (not
necessarily an eigenvalue of $q$), where the $P$ and $M$
are $2\times 2$ matrix operators defined in [Part~I].  It is clear that
$\tilde{x}=[x_2^*,-x_1^*]^T$ satisfies \eqref{eq:dwdt-dwdz}
for $\zeta\rightarrow\zeta^*$, and furthermore, by
cross-elimination, $q$ is a solution of the integrable
equation underlying \eqref{eq:dwdt-dwdz}.

The Darboux theorem is stated as follows.

\begin{theorem}[Darboux transformation]
\label{thm:darboux}
Let $\phi(t, \lambda;q)$ be a known solution of
\eqref{eq:dwdt-dwdz}, and set $\Sigma=S\Gamma S^{-1}$, where
$S=[\phi(t,\lambda;q),\tilde{\phi}(t,\lambda;q)]$ and
$\Gamma=\diag(\lambda,\lambda^*)$.  If $v(t, \mu;q)$
satisfies \eqref{eq:dwdt-dwdz}, then $u(t,\mu;\tilde{q})$
obtained from the Darboux transform
\begin{IEEEeqnarray}{rCl} \label{eq:v-update0}
u(t,\mu;\tilde{q})=\left(\mu I-\Sigma\right)v(t,\mu;q),
\end{IEEEeqnarray}
satisfies \eqref{eq:dwdt-dwdz} as well, for
\begin{IEEEeqnarray}{rCl}
\tilde{q}=q+2j(\lambda^*-\lambda)\frac{\phi_1\phi_2^*}{|\phi_1|^2+|\phi_2|^2}.
\label{eq:qtilde}
\end{IEEEeqnarray}
Furthermore, both $q$ and $\tilde{q}$ satisfy the integrable
equation underlying the system \eqref{eq:dwdt-dwdz}.
\end{theorem}

\begin{IEEEproof}
See Appendix~\ref{app:darboux}.
\end{IEEEproof}

Theorem~\ref{thm:darboux} immediately provides the following
observations.
\begin{enumerate}
\item From $\phi(t,\lambda;q)$ and $v(t,\mu; q)$, we can
obtain $u(t,\mu;\tilde{q})$ according to
\eqref{eq:v-update0}. If $\mu$ is an eigenvalue of $q$, then
$\mu$ is an eigenvalue of $\tilde{q}$ as well. Furthermore,
since $u(t,\mu=\lambda;\tilde{q})\neq 0$, $\lambda$ is also an
eigenvalue of $\tilde{q}$. It follows that the eigenvalues
of $\tilde{q}$ are the eigenvalues of $q$ together with
$\lambda$.
\item $\tilde{q}$ is a new solution of the equation underlying 
\eqref{eq:dwdt-dwdz}, obtained from $q$ according to \eqref{eq:qtilde}, and $u(t,\mu; \tilde{q})$ is one of its
eigenvectors.
\end{enumerate}

\begin{figure*}[t!]
\centerline{
\includegraphics[width=\textwidth]{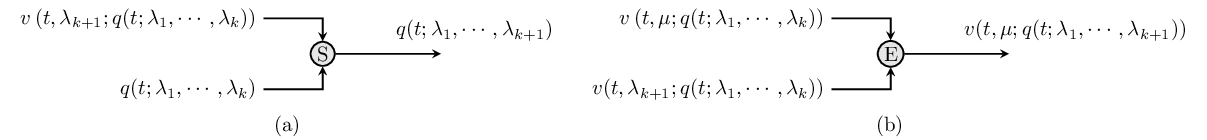}
}
  \caption{Updates in the Darboux transformation: (a) signal update; (b) eigenvector update.}
\label{fig:darboux12}
 \end{figure*}

\begin{figure*}[t!]
\centerline{\includegraphics[width=\textwidth]{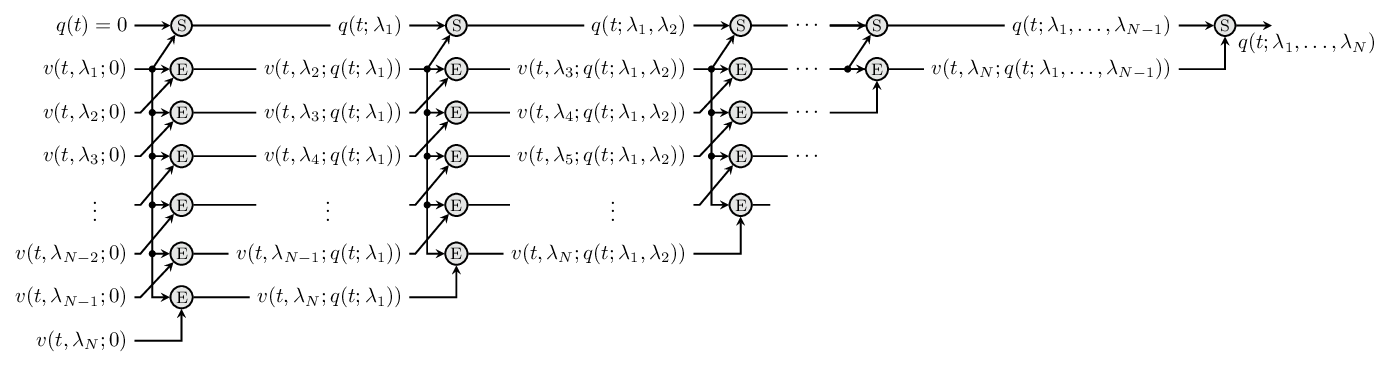}}
\caption{Darboux iterations for the construction of an $N$-soliton.}
\label{fig:darboux3}
 \end{figure*}

These observations suggest a two-step iterative algorithm to
generate $N$-solitons, as illustrated in the
Figs.~\ref{fig:darboux12}--\ref{fig:darboux3}.  Denote a
$k$-soliton pulse with eigenvalues
$\lambda_1,\lambda_2,\ldots,\lambda_k$ by
$q(t;\lambda_1,\lambda_2,\ldots,\lambda_k):=q^{(k)}$.  The
update equations for the recursive Darboux method are given
in Table~\ref{tbl:darboux}.  Note that
$v(t,\lambda_j;q^{(k+1)})$ can also be obtained directly by
solving the Zakharov-Shabat system \eqref{eq:dv-dt-part3}
for $q^{(k+1)}$. It is however more efficient to update the
required eigenvector according to Table~\ref{tbl:darboux}.
The algorithm is initialized from the trivial solution
$q^{(0)}=0$. The initial eigenvectors in
Fig.~\ref{fig:darboux3} are chosen to be the (non-canonical)
eigenvectors $v(t,\lambda_j;0)=[A_je^{-j\lambda_jt},
B_je^{j\lambda_jt}]^T$.  The coefficients $A_j$ and $B_j$
control the spectral amplitudes and the shape of the pulses.
For a single-soliton, $A_j=\exp(j\angle\tilde{q})$ and
$B_j=|\tilde{q}|$.


\begin{remark}
In this paper we mostly use Darboux method for numerical generation of $N$-solitons and discrete spectrum simulations. Hirota method, 
on the other hand, is preferred for the analytical examination of $N$-soliton and insight
into the NFT. The Riemann-Hilbert approach is more general and captures the continuous
spectrum too (though it is sometimes ill-behaved).
\end{remark}

\subsection{Evolution of the Discrete Spectrum}

Recall that the imaginary and real parts of the eigenvalues
correspond, respectively, to soliton amplitude (energy) and
frequency. If the discrete spectrum of the signal lies
completely on the imaginary axis, the $N$-soliton does not travel 
while propagating (with respect to a
traveling observer). The individual
components of an $N$-soliton pulse with frequencies $\lambda_i$ off the $j\omega$ axis
travel in retarded time with speeds proportional to
$\Re\lambda_i$ (frequency).

The manner of $N$-soliton propagation thus depends on the
choice of the eigenvalues. An $N$-soliton signal is
essentially composed of $N$ single-solitons coupled
together, similar to a molecule which groups a number of
atoms.  If the eigenvalues have non-zero distinct real parts, various components
travel at different speeds and eventually, when
$z\rightarrow\infty$, the $N$-soliton decomposes into $N$
separate solitons 
\begin{IEEEeqnarray*}{rCl}
q(t,z)\rightarrow\sum\limits_{i=1}^N \omega_i e^{-j \alpha_i
  t+j(\alpha^2_i-\omega^2_i)z+j\phi_i}\sech(\omega_i(t-2\alpha_i z-t_i)),
\end{IEEEeqnarray*} 
where $\lambda_i=(\alpha_i+j\omega_i)/2$ are eigenvalues and $t_i$ is the time
center. This breakdown of a signal to its individual
components, while best observed in the case of
multi-solitons, is simply a result of group velocity
dispersion and exists for all pulses similarly (including sinc functions).  The extent
of breakdown and shift depends on a variety of factors, such as
the length of the fiber, number of mass points,
fiber dispersion and dispersion-management schemes.  
The effects of pulse broadening
must be carefully considered, especially if dispersion is not managed.

\begin{table*}[t]
\centering
\caption{Update Equations for the Recursive Darboux Method}
\label{tbl:darboux}
\begin{tabular}{|r@{~}c@{~}l|}\hline
\multicolumn{3}{|l|}{Eigenvector update:}\\
\rule{0pt}{5ex}
$ v_1(t,\lambda_j;q^{(k+1)})$&=&$\dfrac{1}{\norm{v(t,\lambda_{k+1};q^{(k)})}^2}\Bigg\{\bigg\{(\lambda_j-\lambda_{k+1})\left|v_1(t,\lambda_{k+1};q^{(k)})\right|^2+(\lambda_j-\lambda^*_{k+1})\left|v_2(t,\lambda_{k+1};q^{(k)})\right|^2\bigg\}
 v_1(t,\lambda_j;q^{(k)})$\\
&&$ +(\lambda_{k+1}^*-\lambda_{k+1})
v_1(t,\lambda_{k+1};q^{(k)})v_2^*(t,\lambda_{k+1};q^{(k)}) v_2(t,\lambda_j;q^{(k)})\Bigg\}$,\\
\rule{0pt}{5ex}$v_2(t,\lambda_j;q^{(k+1)})$&=&
$\dfrac{1}{\norm{v(t,\lambda_{k+1};q^{(k)})}^2}\Bigg\{\bigg\{(\lambda_j-\lambda^*_{k+1})\left|v_1(t,\lambda_{k+1};q^{(k)})\right|^2+(\lambda_j-\lambda_{k+1})\left|v_2(t,\lambda_{k+1};q^{(k)})\right|^2\bigg\}
v_2(t,\lambda_j;q^{(k)})$\\
&&$+(\lambda_{k+1}^*-\lambda_{k+1})v_1^*(t,\lambda_{k+1};q^{(k)})v_2(t,\lambda_{k+1};q^{(k)})v_1(t,\lambda_j;q^{(k)})\Bigg\}$,\\
\multicolumn{3}{|l|}{\rule{0pt}{5ex}for $k=0,\ldots,N-2$ and $j=k+2,\ldots, N$.}\\\hline
\multicolumn{3}{|l|}{Signal update:}\\
$ q^{(k+1)}$&=&$q^{(k)}+2j(\lambda^*_{k+1}-\lambda_{k+1})\dfrac{ v_1(t,\lambda_{k+1};q^{(k)})v_2^*(t,\lambda_{k+1};q^{(k)})}{\norm{v(t,\lambda_{k+1};q^{(k)})}^2}.$\\\hline
\end{tabular}
\end{table*}

\subsection{Demodulating the Discrete Spectrum}

To demodulate a multi-soliton pulse, the eigenproblem
\eqref{eq:dv-dt-part3} needs to be solved. There is limited
work in the mathematical literature concerning the numerical
solution of the Zakharov-Shabat spectral problem
\eqref{eq:dv-dt-part3}. In [Part~II], we have studied
methods by which the nonlinear Fourier transform of a signal
may be computed numerically. In particular, in this paper we use the
layer-peeling and Ablowitz-Ladik methods described in [Part~II] to 
estimate the discrete spectrum.  The reader is
referred to [Part~II] for details.   

\section{Statistics of the Spectral Data}
\label{sec:statistics}

In this section we generalize the deterministic model
considered so far to include the effects of amplified
spontaneous emission (ASE) noise during signal propagation.
We present a method to approximate the statistics of the
spectral data at the receiver.

\begin{remark}
In Section~\ref{sec:wdm} we identified inter-channel interference in
multiuser WDM networks as the intractable factor limiting
the achievable 
rates of the current methods at high launch powers. In
comparison, noise is a weaker form of distortion, and hence
we do not intend to provide here a comprehensive analysis.
\end{remark}

The addition of noise disturbs the vanishing or
periodic boundary conditions usually assumed in the
development of the nonlinear Fourier transform.  One may
therefore question whether the NFT is in fact well defined
in this case.  Fortunately, since the ASE noise power in optical fibers is quite small compared to
the signal power for $\SNR \gg 0~\db$ used in long-haul fiber-optic communications, one can treat noise
as a small perturbation and still safely use the NFT.

Calculation of the exact statistics of the spectral data at
the receiver can be quite cumbersome. This is essentially
because the NLS equation with additive noise, unlike the
noise-free equation, has little or no structure, giving rise
to complicated variational representations for the noise
statistics.  Even if exact expressions could be obtained, it
is unlikely that they would be suitably tractable for data
communications studies. One can, however, approximate these statistics using a perturbation theory, or
simulate them on a computer.  In this paper we follow a
perturbation theory approach.

\begin{remark}
Note that in this paper, we have not included the effects of
fiber loss in our model.  This assumption is justified in
systems using distributed ideal Raman amplification, which
compensates loss but adds an equal amount of noise.
Therefore loss is essentially traded with noise, which is
treated in this section.
\label{rem:loss}
\end{remark}

If noise is added in a lumped fashion, we have the
deterministic NLS equation with random initial data at the
input of each fiber span. In this case, the NFT can be used
without approximation.   

If noise is injected continuously throughout the fiber as a
result of DRA, we have the stochastic NLS equation \eqref{eq:nnls} that includes an additive
space-time noise term. This equation is generally not
integrable\footnote{In a special case, the NLS equation with a certain real-valued multiplicative 
potential can still be integrable
\cite[Appendix D]{ablowitz2003dcn}.}.
 However, we can discretize the fiber into a large number of small fiber
segments and add lumped noise at the end of each segment.
Each such injection of noise acts as a random perturbation
of the initial data at the input of the next segment. The
DRA can thus be approximately treated similar to the lumped
noise case. In this case, NFT is used under such approximation.

In this section, we study the effect of the lumped and distributed noise 
perturbations on the NFDM channel model. We assume that
the noise vanishes, or is negligible, as
$|t|\rightarrow\infty$ and has a finite energy such that the
signal remains absolutely integrable almost surely (NFT assumptions).

\subsection{Perturbation of Eigenvalues}
\subsubsection{Lumped Noise}

The NFT arises in  the spectral analysis of the $L$ operator
\eqref{eq:L-op-part3}. We can easily analyze the
perturbations of the eigenvalues of the $L$ operator to the first order in the noise
level $\epsilon$. 

Let us denote the nonlinear Fourier transform of $q(t)$ in the absence
of noise by $\left(\hat{q}(\lambda), \tilde{q}(\lambda_j)\right)$. As the signal
$q(t)$ at the input of each small segment is perturbed to $q(t)+\epsilon n(t)$ for some small parameter
$\epsilon$ and (normalized) noise process $n(t)$, the (discrete)
eigenvalues and spectral amplitudes deviate slightly from their nominal
values. Separating the signal and noise terms, the perturbed $v$ and
$\lambda$ satisfy
\begin{IEEEeqnarray}{rCl}
(L+\epsilon R)v=\lambda v,\qquad 
R=
\begin{pmatrix}
0 & n \\
-n^* & 0
\end{pmatrix},
\label{eq:L+dL}
\end{IEEEeqnarray}
where $R$ is the matrix containing the noise.  The study of
the nonlinear Fourier transform in the presence of (small)
input noise is thus a perturbation theory of the non-self-adjoint
operator $L+\epsilon R$.

Perturbation theory of Hermitian operators is well-studied (\eg, in quantum mechanics).  The Zakharov-Shabat
operator in \eqref{eq:L+dL} is however non-self-adjoint.
Unfortunately most useful properties of self-adjoint
operators (in particular, the existence of a complete
orthonormal basis from eigenvectors) do not carry over
to non-self-adjoint operators.  For either type of operator,
deterministic perturbation analysis already exists in the
literature \cite{karpman1981pat, karpman1977apt,
kaup1976ape, konotop1994nrw}.  These results, however, are
non-stochastic and the distribution of the scattering data
is still lacking. A very interesting work is
\cite{moustakas2008nse} in which authors calculate the
distribution of the spectral data for the special case in
which the channel is noise-free and the input is a white
Gaussian stochastic process. There is also much work 
pertaining to the statistics of the parameters of a single-soliton; 
see \eg , \cite{falkovich2004nongaussian} and references 
therein.

For the non-self adjoint operators $L$, the orthogonality
that we require is between the space of left and right
eigenvectors of $L$ associated with distinct eigenvalues;
that is to say, between eigenvectors of $L$ associated with
$\lambda$ and eigenvectors of the adjoint operator $L^*$
associated with $\mu\neq \lambda^*$. Let us equip the space
of eigenvectors with the usual $L^2$ inner product
\begin{IEEEeqnarray*}{rCl}
\inner{u}{v}=\int\limits_{-\infty}^{\infty}\left(u_1v_1^*+u_2v_2^*\right)dt.
\end{IEEEeqnarray*}
It can be verified that the operator $\Sigma_3 L$ is
self-adjoint, \ie,
$\inner{u}{\Sigma_3Lv}=\inner{\Sigma_3Lu}{v}$, where
$\Sigma_3=\diag(1,-1)$ is the Pauli matrix.

We use a small noise approximation, expanding unknown variables in noise
level $\epsilon$ as
\begin{IEEEeqnarray}{rCl}
v(t)&=&v^{(0)}(t)+\epsilon v^{(1)}(t)+\epsilon^2 v^{(2)}(t)+\cdots ,
\IEEEyesnumber\IEEEyessubnumber
\label{eq:v-series}
\\
\lambda&=&\lambda^{(0)}+\epsilon \lambda^{(1)}+\epsilon^2\lambda^{(2)}+\cdots.
\IEEEyessubnumber
\label{eq:lambda-series}
\end{IEEEeqnarray}
We assume these variables are analytic functions of
$\epsilon$ so that the above series are convergent.
Plugging \eqref{eq:v-series}--\eqref{eq:lambda-series} into
\eqref{eq:L+dL} and equating like powers of $\epsilon$, we
obtain
\begin{IEEEeqnarray}{rCl}
Lv^{(0)}&=&\lambda^{(0)} v^{(0)} \label{eq:per0},\\
(L-\lambda^{(0)})v^{(1)}&=&-(R-\lambda^{(1)})v^{(0)} \nonumber\label{eq:per1},\\
(L-\lambda^{(0)})v^{(2)}&=&-(R-\lambda^{(1)})v^{(1)}+\lambda^{(2)} v^{(0)} \nonumber\label{eq:per2},
\end{IEEEeqnarray}
and so on. The first term implies that $v^{(0)}$ and $\lambda^{(0)}$ are eigenvalue
and eigenvector of the (nominal) operator $L$. To eliminate $v_1$ from
the second equation, we take the inner product on both sides of
\eqref{eq:per0} with some vector $u$; the left hand side of the
resulting expression is 
\begin{IEEEeqnarray}{rCl}
  \inner{u}{(L-\lambda^{(0)})v^{(1)}}&=&\nonumber
  \inner{(L-\lambda^{(0)})^*u}{v^{(1)}}\nonumber\\
&=&  \inner{(L^*-\lambda^{(0)*})u}{v^{(1)}}.
\label{eq:kill-v1}
\end{IEEEeqnarray}
To have the right-hand side of \eqref{eq:kill-v1} vanish, we
can choose $u$ to be an eigenvector of the adjoint operator
$L^*$ associated with an eigenvalue $\mu=\lambda^{(0)*}$,
\ie, $(L^*-\lambda^{(0)*})u=0$.  Since $L^*(q)=L(-q)$, if
$Lv=\lambda^{(0)} v$, it can be verified that
$L^*u=\lambda^{(0)}u$ for $u=[v_1,-v_2]=\Sigma_3 v$.
Setting $u=u^{(0)}=\Sigma_3 v^{(0)}(t,\lambda^*)$,
\begin{IEEEeqnarray*}{rCl}
\lambda^{(1)}=\frac{\inner{u^{(0)}}{Rv^{(0)}}}{\inner{u^{(0)}}{ v^{(0)}}}.
\end{IEEEeqnarray*}
Using similar calculations we obtain $\lambda^{(2)}$ 
\begin{IEEEeqnarray*}{rCl}
\lambda^{(2)}&=&\frac{\inner{u^{(0)}}{Rv^{(1)}}}{\inner{u^{(0)}}{v^{(0)}}}-
\lambda^{(1)}\frac{\inner{u^{(0)}}{v^{(1)}}}{\inner{u^{(0)}}{v^{(0)}}},
\end{IEEEeqnarray*}
and so on.

To summarize, the fluctuations of the $n^{\text{th}}$ eigenvalue
$\lambda_n$ with nominal eigenvector $v_n$ is given by
\begin{IEEEeqnarray}{rCl}
\hat{\lambda}_n=\lambda_n+\epsilon\frac{\inner{u_{n}}{Rv_n}}{\inner{u_n}{v_n}}+O(\epsilon^2),\quad n=1,2,\ldots,N,
\label{eq:lambda-fluc-lumped}
\end{IEEEeqnarray}
where $u_n=\Sigma_3 v_n$ and `$^\wedge$' denotes the eigenvalue after noise addition. It follows that the perturbation of the eigenvalues is
distributed, to the first order, according to a zero-mean
complex Gaussian distribution.

Continuing this approach to find higher-order fluctuations
of eigenvectors, $v^{(k)}$, $k\geq 1$, is not
straightforward because the underlying operator is not
self-adjoint.

\subsubsection{Distributed Noise}

Consider now the perturbed NLS equation
\begin{IEEEeqnarray}{rCl}
\label{eq:snls}
jq_z=q_{tt}+2|q|^2q+\epsilon n(t,z),
\end{IEEEeqnarray}
where $\epsilon$ is a small parameter (noise power) and
the normalized noise term $n(t,z)$ represents the combined effects of the signal loss
and the distributed noise. 
 
Let us represent \eqref{eq:snls} with the same $L$ and $M$
of the noise-free equation and now let $\lambda$ vary with
$z$. The equality of mixed derivatives $v_{tz}=v_{zt}$
gives
\begin{IEEEeqnarray*}{rCl}
\begin{pmatrix}
-j\lambda_z & q_z+jq_{tt}+2j|q|^2q \\
-q^*_z+jq_{tt}^{*}+2j|q|^2q^* & j\lambda_z
\end{pmatrix}v=0. 
\end{IEEEeqnarray*}
This, upon re-arranging and using \eqref{eq:snls}, simplifies to 
\begin{IEEEeqnarray*}{rCl}
\lambda_z v=\epsilon \bar{R}v,\quad  \bar{R}=-R.
\end{IEEEeqnarray*}

Note that, as before, we do not have $v(t,z)$ a priori
because, according to \eqref{eq:L-op-part3}, it depends on
the noisy signal $q(t,z)$ and $\lambda(z)$, both of which
are unknown.  However, if the noise level $\epsilon$ is
small, we can expand $v(t,z)$ and $\lambda$ in powers of
$\epsilon$ as in \eqref{eq:v-series} and
\eqref{eq:lambda-series} to obtain $(\lambda^{(0)})_z=0$ and
$\bar{R}v^{(0)}=(\lambda^{(1)})_zv^{(0)}$ (\ie,
$(\lambda^{(1)})_z$ appears as a time-independent eigenvalue
of $\bar{R}(t)$). Taking the inner-product with
$u^{(0)}=\Sigma_3 v^{(0)}(t,\lambda^*)$ on both sides of
$\bar{R}v^{(0)}=(\lambda^{(1)})_zv^{(0)}$, we obtain the
first-order variation of eigenvalues
 \begin{IEEEeqnarray}{rCl}
(\lambda_1)_z =\frac{\inner{u^{(0)}}{\bar{R}v^{(0)}}}{\inner{u^{(0)}}{v^{(0)}}}.
\label{eq:lambdaz}
\end{IEEEeqnarray}

It follows from \eqref{eq:lambdaz} that the distribution of
the deviation of the eigenvalues is approximately a
zero-mean conditionally Gaussian random variable. The
variance of this random variable is signal-dependent, and
although eigenvectors of an $N$-soliton can be represented
as a series from Darboux transform, it is best calculated
numerically if $N\geq 2$.

\begin{example}
  \label{ex:one-eig-pert}
Consider the single-soliton of
Example~\ref{ex:single-soliton}.  It can be verified that
$\inner{\Sigma_3 v(t,\lambda^*)}{v(t,\lambda)}$
$=-\frac{\omega}{2}(1+\frac{\omega^2}{|\tilde{q}|^2})=-\frac{1}{\omega}$,
where we assumed $|\tilde{q}|=\omega$ so that the soliton is
centered at $t_0=0$. Furthermore,
\begin{IEEEeqnarray*}{rCl}
&&\inner{\Sigma_3v(t,\lambda^*)}{\bar{R}v(t,\lambda)}\\
&=&-\int\left(v_1(t,\lambda^*)v_2^*(t,\lambda)n^*+v_2(t,\lambda^*)v_1^*(t,\lambda)n\right)\der t\nonumber\\
&=&-\int \frac{1}{4}\sech^2(\omega
t)\left(v_2(t,\lambda^*)-v_2^*(t,\lambda)\right)\Re\bar{n}\der t\\
&&-j\int
\frac{1}{4}\sech^2(\omega t)(v_2(t,\lambda^*)+v_2^*(t,\lambda))\Im{\bar{n}}\der t\nonumber\\
&=&\frac{1}{2}\int \sech(\omega t)\tanh(\omega t)\Re\bar{n}\der t-
\frac{j}{2}\int \sech(\omega t)\Im{\bar{n}}\der t,
\end{IEEEeqnarray*}
where $\bar{n}=n\exp(j\Phi)$ has the same statistics as $n$.
It follows that
\begin{IEEEeqnarray}{rCl}
  \alpha_z&=&-\int \sech(\tau)\tanh(\tau)z_1(\tau,z)\der \tau
  \IEEEyesnumber \IEEEyessubnumber\label{eq:omegaz},\\
  \omega_z&=&\int \sech(\tau)z_2(\tau,z)\der \tau,
\IEEEyessubnumber
\label{eq:alphaz}
\end{IEEEeqnarray}
where $z_1=\Re z$, $z_2=\Im z$ and $z(t,z)=\bar n(t/\omega,z)$, \ie, $z_1$ and $z_2$ are independent Gaussian processes each with total power $\omega W z\sigma_0^2$.
The Gordon-Haus effect can be observed from the $\alpha_z$
equation. Note that in fiber optics noise is added to the
signal, \ie, $n$ in this subsection should be replaced
with $j n$. 
\end{example}  

\begin{remark}
Note that the higher-order terms $\lambda^{(i)}$ in expansion
\eqref{eq:lambda-fluc-lumped} are signal-dependent 
(even though they are normalized by $\langle u_n, v_n\rangle$). For instance, Example~\ref{ex:one-eig-pert}, and as well as
Fig.~\ref{fig:constellation} (b), show that noise variance grows with signal. In this case 
higher-order terms need to be included as well for precision. The perturbation
approach given in this section takes into account only the
first term in the expansion, and as a result mostly
describes the bulk of the
distribution in the small noise limit. Note, however, that the first term is also signal-dependent. Thus the effect of the noise growth with
signal is accounted for in the above analysis.
\end{remark}

\subsection{Perturbation of Spectral Amplitudes}
Using a similar perturbation approach, we can study the influence of
noise on spectral amplitudes as well.

As reported in [Part~II], in the Riemann-Hilbert approach the discrete spectral amplitudes are chaotic
even when noise is zero. We thus here consider first-order fluctuation of 
continuous spectral amplitudes, under the lumped noise model. 

Continuous spectral amplitudes are obtained by solving
the following Riccati equation [Part~I]
\begin{IEEEeqnarray}{rCl}
\frac{\der y(t,\lambda)}{\der t}=-\left(\bar{q}(t,\lambda)+\epsilon n(t)\right)y^2(t,\lambda)-
\left(\bar{q}^*(t,\lambda)+\epsilon  n^*(t)\right),\nonumber\\
\label{eq:riccati}
\end{IEEEeqnarray}
where $y(-\infty,\lambda)=0$ and
\[
\bar{q}(t,\lambda)=q(t)\exp(2j\lambda t),\quad
\hat{q}(\lambda)=\lim_{t\rightarrow\infty}y(t,\lambda).
\]
One can write a Fokker-Planck equation for the probability distribution of
$y(t,\lambda)$ in \eqref{eq:riccati} \cite{yousefi2010fokker,yousefi2011opc}. If the signal $q(t)$ is known,
\eg , $q(t)=0$, the resulting equation
can be solved. In general, however, we can expand
\begin{IEEEeqnarray*}{rCl}
  y(t,\lambda)&=&y_0(t,\lambda)+\epsilon y_1(t,\lambda)+\epsilon^2
  y_2(t,\lambda)+\cdots,\\
\hat{q}(\lambda)&=&\hat{q}_0(\lambda)+\epsilon \hat{q}_1(\lambda)+\epsilon^2 \hat{q}(\lambda)+\cdots,
\end{IEEEeqnarray*}
and equate like powers of $\epsilon$. We obtain that
$\hat{q}_0(\lambda)$ (and its corresponding $y_0(t,\lambda)$)
is the spectral amplitude when noise is zero (assuming
$\bar{q}(t,\lambda)\approx \bar{q}(t,\lambda_0)$). Define
\begin{IEEEeqnarray*}{rCl}
  G(t,\lambda)=2\int\limits_{-\infty}^{t}\bar{q}(\tau,\lambda)y_0(\tau,\lambda)\der\tau,
\end{IEEEeqnarray*}
and $\hat{n}(t,\lambda)=-\left(n(t)y_0^2(t,\lambda)+n^*(t)\right)$. Then 
\begin{IEEEeqnarray*}{rCl}
  \hat{q}_1(\lambda)=e^{-G(\infty,\lambda)}\int_{-\infty}^{\infty}\hat{n}(\tau,\lambda)
e^{G(\tau,\lambda)}\der\tau.
\end{IEEEeqnarray*}
This is a conditionally Gaussian random variable (for each $\lambda$)
with variance
\begin{IEEEeqnarray*}{rCl}
  \E\left|\hat{q}_1(\lambda)\right|^2=\sigma_0^2e^{-2\Re
    G(\infty,\lambda)}\int_{-\infty}^{\infty}
\left(|y_0(\tau,\lambda)|^4+1\right)
e^{2\Re G(\tau,\lambda)}\der \tau,
\end{IEEEeqnarray*}
where we assumed that noise is delta-correlated. If for some signals,
$\E|\hat{q}_1(\lambda)|^2$ is unbounded, the above perturbation expansion
fails and a slow-scale variable $T=\epsilon t$ needs to be introduced.

In summary, in this section we showed that a simple first-order perturbation
analysis, though inaccurate, gives insights into the nature of the statistics in the nonlinear spectral 
domain. In the next section, we discuss the impact of the noise
on the achievable spectral efficiencies.   

\section{Some Achievable Spectral Efficiencies Using the NFT}
\label{sec:examples}

We now turn to a numerical study of data modulation in the nonlinear Fourier domain, 
providing some simulation results
and examples of achievable spectral efficiencies. First, in Sections~\ref{sec:1-sol} to \ref{sec:N-sol}, we
set the continuous spectrum to zero and modulate the discrete spectrum. This special case corresponds to an $N$-soliton
data transmission system. Then, in Section~\ref{sec:contin}, we set the discrete spectrum to zero and modulate 
the continuous spectrum.

For the discrete spectrum modulation, we begin with a classical on-off keying soliton
transmission ($N=1$), in which, in any symbol period $T_s$, either
zero or a fundamental soliton is sent. We then increase $N$ and the spectral
efficiency by considering an
$(N\geq 2)$-soliton system, occupying the same time interval as
the on-off keyed soliton system, and maintaining the same
bandwidth requirements. The location of the eigenvalues and
the values of the discrete spectral amplitudes can be
jointly modulated for this purpose. We shall see that the
effective useful region in the upper half-plane to exploit
the potential of discrete eigenvalue modulation is limited
by a variety of factors. 

Modulating the nonlinear spectrum generates pulses with
variable width, power, and bandwidth. We take the average
time, average power, and the maximum bandwidth to properly
convert bits/symbol to bits/s/Hz.  As a first step to improve
upon the on-off keying solitons, first we continuously
modulate one eigenvalue in a given region (\ie, a classical
soliton but with varying amplitude, width and phase). We next
consider multi-soliton systems with a number of
constellations on eigenvalues and discrete spectral
amplitudes.

Throughout this section, we consider a 2000~km single-mode
single-channel optical link in which fiber loss is perfectly
compensated in a distributed manner using Raman
amplification.  Fiber parameters are given in
Table~\ref{tbl:fiberparam}.  Dispersion compensation is not
applied, as it is an advantage of the NFT approach that no
optical dispersion management or nonlinearity compensation is
required. We let pulses interact naturally, as atoms in a
molecule, and perform signal processing at the receiver on these groups. The
method however works for dispersion managed fibers as well,
and in general with operations that do not change
integrability.

\subsection{Spectral Efficiency of 1-Soliton Systems}
\label{sec:1-sol}

\begin{figure*}[t!]
\centerline{\includegraphics[scale=0.8]{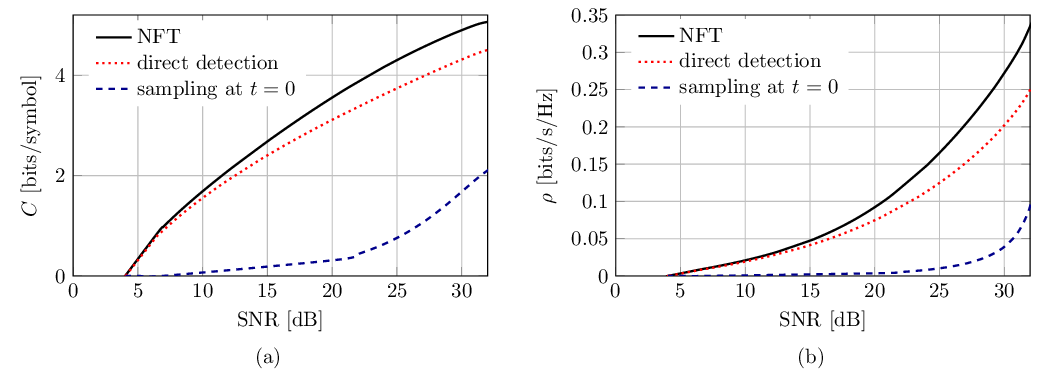}}
  \caption{ (a) Capacity (bits/symbol) and, (b) spectral efficiency
    (\rm{bits/s/Hz}), of soliton systems using direct detection,
    sampling at $t=0$, and the NFT. 
}
\label{fig:soliton-cap}
\end{figure*}

Traditional soliton transmission systems typically do not have high
spectral efficiencies. This is because the amplitude and the
width of a single-soliton are inversely related, and hence
they require a lot of time or bandwidth per degree-of-freedom provided.  
Errors in a soliton transmission system
occur either because of the Gordon-Haus timing jitter effect
(which is the primary source of the errors, if not managed)
\cite{mollenauer2006sof} or amplitude (energy) fluctuations.
It follows from Galilean invariance \cite{ablowitz2006sai}
that the Gordon-Haus effect exists for all kinds of pulses
to the same extent and is not specific to solitons. This
classical effect can be reduced with the help of suitably
designed filters.   We do not treat Gordon-Haus jitter
analytically; however, system simulations naturally include this
effect.

Let us first consider a classical soliton system with only one
eigenvalue $\lambda=(\alpha+j\omega)/2$. The joint density
$f_{A,\Omega}(\alpha,\omega)$ at any fixed distance $z$ can be obtained from
\eqref{eq:alphaz}--\eqref{eq:omegaz} (or by
extracting the dynamics of $\alpha$ and $\omega$ from the
stochastic NLS equation, resulting in a pair of coupled
stochastic ordinary differential equations).

Note that a soliton of the deterministic NLS equation launched into a system
described by the stochastic NLS equation would, of course,
have a growing continuous spectrum too.  In addition, there
would be a small chance of creating additional solitons
out of noise at some distance, or the soliton spectrum might
collapse into the real axis in the $\lambda$ plane. All these effects 
are negligible if noise is small enough, $2Wz\sigma^2 \ll E(0)$, and the propagation distance is not
exceedingly long. Thus, at a length scale determined by $2Wz\sigma^2\ll E(0)$, we can still 
think of the noisy signal as a soliton with re-modulated parameters.

Multiplying the stochastic NLS equation \eqref{eq:snls} by
$q^*$, subtracting from its conjugate, integrating over
time, and using integration by parts in the dispersion
term, we get
\begin{IEEEeqnarray*}{rCl}
\frac{\partial E}{\partial
  z}=2\Im\int\limits_{-\infty}^{\infty}q(\tau,z)Z(\tau,z)\der\tau,
\end{IEEEeqnarray*}
where $E(z)=\int|q(\tau,z)|^2\der\tau$ is the energy, and
$Z=-\epsilon n^*$ is a noise process similar to $\epsilon n$. Replacing $q(t,z)\rightarrow q(t,0)$ in the small
noise limit, we obtain that energy
fluctuation is a signal-dependent conditionally Gaussian random variable
$E(z)\approx \normalr{E(0)}{\sigma^2E(0)}$ ($\sigma^2=2Wz\sigma_0^2/P$). Ignoring the energy of the continuous spectrum 
in the small noise limit $\sigma^2\ll 1$, we have
$E\approx 2\omega$ and therefore
\begin{IEEEeqnarray}{rCl}
\omega(z)=\omega(0)+\sigma\sqrt{\frac{\omega(0)}{2}}\tilde{n},\quad \omega(0) \gg \sigma^2,
\label{eq:omegaz-omega0}
\end{IEEEeqnarray}
where $\tilde{n}\sim\normalr{0}{1}$. The conditional probability density function (PDF) is
\begin{IEEEeqnarray}{rCl}
f_{\Omega|\Omega_0}(\omega|\omega_0)=\frac{1}{\sqrt{\pi\sigma^2\omega_0}}e^{-\frac{(\omega-\omega_0)^2}{\sigma^2
    \omega_0}},\quad
\omega_0=\omega(0),
\label{eq:omega-pdf}
\end{IEEEeqnarray}
and the PDF of $r=\sqrt{\omega}$ given $r_0=\sqrt{\omega_0}$,
$\omega,\omega_0\geq 0$, is approximately a Rician distribution
\begin{IEEEeqnarray*}{rCl}
f_{R|R_0}(r |r_0)&=&\frac{1}{\sqrt{\pi \sigma^2} }e^{-\frac{(r-r_0)^2}{\sigma^2}}\\
&\approx&
\frac{2r}{\sigma^2}e^{-\frac{r^2+r_0^2}{\sigma^2}}I_0(\frac{2rr_0}{\sigma^2}),
\quad r,r_0\gg \sigma,
\label{eq:sqrt-omega-pdf}
\end{IEEEeqnarray*}
which is signal-independent in the high \SNR\ regime.

In \cite{yousefi2011opc} we have shown that a half-Gaussian density
\begin{IEEEeqnarray*}{rCl}
f_{\Omega_0}(\omega_0)=\frac{2}{\sqrt{2\pi \const{P}} }e^{-\frac{\omega^2_0}{2\const{P}}},\quad \omega_0\geq 0,
\end{IEEEeqnarray*}
gives the asymptotic capacity for \eqref{eq:omega-pdf}
\begin{IEEEeqnarray*}{rCl}
C \sim \frac{1}{2}\log (\SNR),
\end{IEEEeqnarray*}
where $\SNR=\frac{\const{P}}{\sigma^2}$. 

Translating capacity in bits/symbol to spectral efficiency in
\rm{bits/s/Hz} depends on the receiver architecture.  Assuming that the
receiver is able to decode pulses with variable widths, the spectral
efficiency $\rho(\const{P})$ is obtained by
\begin{IEEEeqnarray}{rCl}
\rho(\const{P})&=&\max\limits_{\substack{f(\omega_0)\\\omega_0\in
    S}}\frac{1}{\textrm{BW}(S) \E T(\omega_0) }I(\omega_0;\omega),\label{eq:rho-prob}\\
\E P(\omega_0)&\leq&\const{P},
\nonumber
\end{IEEEeqnarray}
where $T(\omega_0)$ and $P(\omega_0)$ are the width and the power of a
single-soliton with amplitude $\omega_0$ and $\textrm{BW}(S)=\max\limits_{\omega_0\in S}
\textrm{BW}(\omega_0)$ is the maximum passband bandwidth that the signal set
 requires for transmission. For a one soliton signal,
we have approximately 
\begin{IEEEeqnarray}{rCl}
T(\omega_0)=\frac{7}{\omega_0}, \quad P(\omega_0)=\frac{\omega_0^2}{6.2},\quad \textrm{BW}(\omega_0)=0.95\omega_0,
\label{eq:T-P-BW}
\IEEEeqnarraynumspace
\end{IEEEeqnarray} 
where the width $T(\omega_0)$ includes a guard time---four
times the full width at half maximum power (FWHM)---so as to
minimize the intra-channel interactions.
 
Using \eqref{eq:T-P-BW}, the maximum spectral efficiency of
a baseline on-off keying system is obtained to be about
$\rho\approx 0.15$ $\rm{bits/s/Hz}$ at the average power
$\const{P}_0=0.16~\rm mW$. Note that the per unit cost capacity problem
\eqref{eq:rho-prob} is non-convex and hence finding the
global optimum may prove to be challenging. Here we simply
optimize mutual information and scale it by $\text{BW}(S)\times \E T(\omega_0)$ evaluated at the
mutual-information-maximizing input distribution. 

\begin{figure}[t!]
\centerline{\includegraphics[width=\columnwidth]{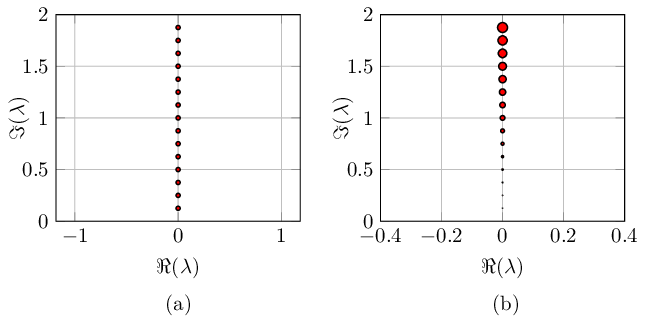}}
\caption{
(a) Eigenvalue constellation at the transmitter. (b)
    Noise balls at the receiver in the NFT approach. The
    signal-dependency of the noise balls can be analytically seen \eg,
    through \eqref{eq:omegaz-omega0}. The radius of each circle is the
  standard deviation of the received eigenvalue.}
\label{fig:constellation}
 \end{figure}

Fig.~\ref{fig:soliton-cap} shows the achievable rate and the
spectral efficiency of a 1-soliton system with amplitude
modulation using various detection methods. Note that since
we do not solve the optimization problem
\eqref{eq:rho-prob}, the spectral efficiencies shown in the
Fig.~\ref{fig:soliton-cap}(b) are only lower bounds on the
actual achievable values. Figs.~\ref{fig:constellation}(a)--(b)
show the constellation at the transmitter and the ``noise
balls'' (of radius equal to one standard deviation of the
distance to the transmitted point) accumulated over 30,000
simulation trials at the receiver. The actual number of
signal levels is 64 in the simulations. Calculation of the
approximate rate is performed using the Arimoto-Blahut
algorithm and is confirmed by numerical interior point
optimization.

Note that, as is clear from
Figs.~\ref{fig:soliton-cap}(a)--(b), sampling
signals at $t=0$ is clearly a bad idea; it is shown here just to see
the effects of the timing jitter on $1-$soliton systems. 

\subsection{Spectral Efficiency of 2-Soliton Systems}
\label{sec:2-sol}
To illustrate how the NFT method works, we start off with
two simple examples.  These two examples are intended to
explain the details of transmission and detection using the
NFT, but they have not been optimized for performance.

\subsubsection{Modulating Eigenvalues}
\label{sec:example1}

Consider the following signal set with 4 elements:
\begin{IEEEeqnarray}{rCl}
&S_1&\quad : \quad 0, \nonumber \\ 
&S_2&\quad:\quad \tilde{q}(0.5j)=1,\nonumber \\
&S_3&\quad:\quad \tilde{q}(0.25j)=0.5, \nonumber\\
&S_4&\quad : \quad \tilde{q}(0.25j,0.5j)=(1,1).
\label{eq:exp1-signals}
\end{IEEEeqnarray}
Table \ref{tbl:signal-params} shows the energy, duration,
power and the bandwidth of these signals.

\begin{table}[b]
\centerline{
\begin{threeparttable}[flushleft]
\caption{Parameters of the signal set in \eqref{eq:exp1-signals}\tnote{1}.}
\renewcommand{\tabcolsep}{3pt}
\begin{tabular}{c|c|c|c|c|c}
signal & energy & duration FWHM & 99\% duration& power & bandwidth \\
$S_1$ & 0 & $T_0$ & $T_1$ & 0 & $W_0$ \\ 
$S_2$    &      $E_0$ & $T_0$&$T_1$ &$P_0$& $W_0$\\
$S_3$ &       0.5 $E_0$ & 2$T_0$& $2T_1$&0.25$P_0$&0.5$W_0$ \\
$S_4$ & 1.5 $E_0$ & 4.25$T_0$& $2.58T_1$ &0.58$P_0$& 0.5$W_0$\\
\end{tabular}
\begin{tablenotes}
\footnotesize
\item [1]
Here
  $E_0=2$, $T_0=1.763$ at
FWHM power, $T_1=5.2637$ (99\% energy),
$P_0=0.38$ and $W_0=0.5714$. The normalization factors in the NLS equation are $T_n=25.246~\rm ps$ and $P_n=0.5~\rm mW$ 
at dispersion $0.5~\rm ps/(nm-km)$ and $\gamma=1.27~\rm{W^{-1}km^{-1}}$.
\end{tablenotes}
\label{tbl:signal-params}
\end{threeparttable}
}
\end{table}

We compare this with a standard on-off keying (OOK) soliton
transmission system, consisting of $S_1$ and $S_2$.  From
the signal parameters given in
Table~\ref{tbl:signal-params}, it follows that the OOK system
provides about $\rho_0=0.33$~bits/s/Hz spectral efficiency
at $\const{P}_0=0.1876$~mW and $R_0=7.42~\rm Gbits/s$
data rate.  Note that the noise level is so small compared
to the imaginary part of the eigenvalues that this scheme
essentially achieves a transmission rate of 2 bits/symbol.

The full constellation defined in (\ref{eq:exp1-signals})
has average power $\const{P}=0.46P_0$ and average time duration
$T=1.65T_1$, where $P_0$ and $T_1$ are the power and the time
duration of the fundamental soliton.  The new signal set
therefore provides a spectral efficiency of about
$\rho=\log 4/(1.65T_1 W_0)=1.2121\times \rho_0$~bits/s/Hz
and operates at $R=1.2121\times R_0$ for about the same
average power ($\const{P}=0.1748$~mW). Note that without $S_4$ the
average power would be higher and in addition the
improvement in the spectral efficiency would be slightly
smaller compared to the on-off keying system.  Signal $S_4$
is the new signal (a 2-soliton) that goes beyond
conventional pulse shapes.  Such signals do not cost much in
terms of $\textnormal{time}\times\textnormal{maximum
bandwidth}$ product, while they add additional elements to
the signal set. These additional signals can generally be
best decoded with the help of the nonlinear Fourier
transform. 

In this example, the receiver needs to estimate the
pulse-duration.  This can be done in many ways, \eg, using
the NFT computations already performed: zeros of the signal
in time can be detected when $[v_1^2e^{j\lambda t}, v_2^2
e^{-j\lambda t}]$ reaches a constant value in steady state.
This can be checked at times $t=T_1$, $t=2T_1$ and
$t=2.58T_1$. If one of the signals is zero at the end of
another signal, one can monitor the energy of the continuous
spectrum to make sure that it is small. If the symbol duration
is fixed to be the maximum $2.58T_1$, the addition of $S_3$
and $S_4$ increases both time interval and cardinality of
signal set such that the spectral efficiency and data rate
remain constant ($\log(6)/2.58$), while operating at
$77\%$ of the on-off keying signal power. 

Since solitons with purely imaginary eigenvalues do not
suffer from major temporal or spectral broadening, spectral
efficiencies at the fiber output are essentially the same as
those at the input to the fiber.

\begin{figure*}
\centering
\includegraphics{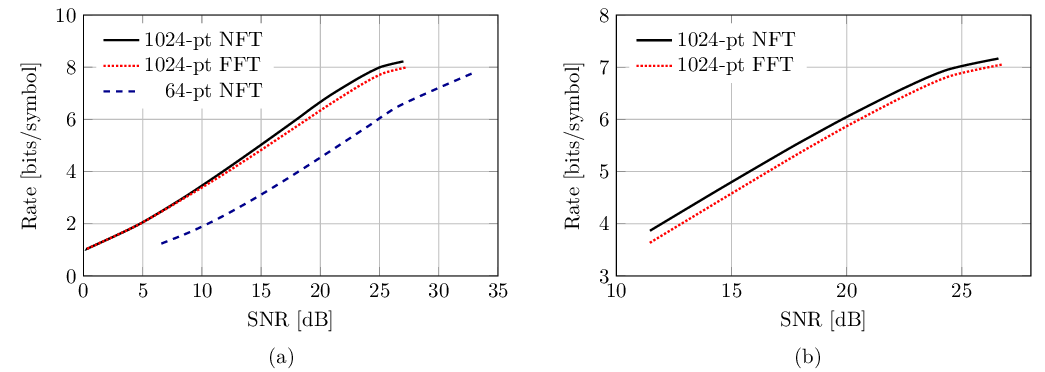}
  \caption{Achievable rates in (a) a single-channel, and (b) a WDM optical fiber
    system using the nonlinear Fourier transform and backpropagation. The
\SNR\ is calculated at the system bandwidth and can be adjusted to represent the \emph{optical} signal-to-noise ratio. Note that since we have
used raised-cosine functions with 50\% excess bandwidth, $\rho=\frac{2}{3}C$.}
\label{fig:capacity-contin}
 \end{figure*}

\subsubsection{Modulating Eigenvalues and Spectral Amplitudes}

We can improve upon the previous example by modulating the
spectral amplitudes as well. Consider the following signal set  
\begin{IEEEeqnarray*}{rCl}
&S_1& \quad : \quad 0, \nonumber \\ 
&S_2-S_5&\quad:\quad \tilde{q}(0.5j)=\tilde{q}_1,\nonumber \\
 &S_6-S_9&\quad:\quad \tilde{q}(0.25j)=\tilde{q}_2, \nonumber\\
&S_{10}-S_{16}&\quad : \quad \tilde{q}(0.25j,0.5j)=(\tilde{q}_3,\tilde{q}_4).
\label{eq:exp2-signals}
\end{IEEEeqnarray*}
We make a 3-ary constellation on
$\tilde{q}_i\in\left\{0.5, 1, 1.5\right\}$. This creates a
signal set with 16 elements. Here pulses are extended to
$3T_1$ time duration. The resulting constellation has
average power $\const{P}=1.06P_{0}$ and average time duration
$T=2.236T_1$, where $P_0$ and $T_1$ are the power and the
symbol-duration
of the benchmark on-off keying system. Therefore
the new signal set provides about $\rho=\log 16/(2.236T_1
W_0)=1.79\times \rho_0$ \rm{bits/s/Hz} and operates at
$R=1.79 \times R_0$ for about the same average power. If we
fix symbol durations to be the maximum $3T_1$, then the
improvement is $\rho=2.2\rho_0=0.73$ \rm{bits/s/Hz}, at
$80\%$ of the average power.

Again, since the real part of the eigenvalues is not
modulated, signals do not suffer from major temporal or
spectral broadening.

\begin{remark}
Note that modulating the eigenvalues includes only the
amplitude information (similarly to $M$-ary
frequency-shift-keying).  To excite the other half of the
degrees-of-freedom representing the phase, discrete spectral
amplitudes should also be considered. While
$|\tilde{q}(\lambda_j)|$ may be noisy, the phase $\angle
\tilde{q}(\lambda_j)$ or a function of
$\left\{\tilde{q}(\lambda_j)\right\}_{j=1}^{j=N}$ can be
investigated for this purpose.  
\end{remark}

\subsection{Spectral Efficiency of $N$-Soliton Systems, $N\geq 3$}
\label{sec:N-sol}
To achieve greater spectral efficiencies, a dense
constellation in the upper half complex plane needs to be
considered. A spectral constellation with $n$ possible
eigenvalues in $\Complex^+$ (from which $k$ eigenvalues are
chosen, $0 \leq k \leq n$) and $m$ possible values for
spectral amplitudes provides up to
\begin{IEEEeqnarray*}{rCl}
\log\left(\sum\limits_{k=0}^{n}\binom{n}{k}m^k\right)=n\log\left(m+1\right),
\end{IEEEeqnarray*}
bits per symbol (fewer if a subset is chosen).  One can continue the
approach presented in the previous examples by increasing
$n$ and $m$. The receiver architecture presented in [Part~I] 
is fairly simple and is able to decode NFT signals
rather efficiently.

Some choices of spectral parameters may translate to pulses having a
large peak to average power ratio, large
(99\%) bandwidth, or large (99\%) time duration at $z=0$ or during their
propagation. The signal set should thus be expurgated to
avoid such undesirable signals.  We have not yet found rules
for modulating the spectrum so that such undesirable signals
are not generated.  For the small examples given here, we
can check pulse properties directly; however, appropriate
design criteria for the spectral data (particularly the
discrete spectral amplitudes) should be developed.

In this simulation, we assume a constellation with 30 points
uniformly chosen in the interval $0\leq \lambda \leq 2$ on
the imaginary axis and create all $N$-solitons, $1\leq N\leq
6$. We then prune signals with undesirable bandwidth or
duration from this large signal set. The remaining
multi-solitons are used as carriers of data in the
fiber system described at the beginning of Section~\ref{sec:examples}. Here a spectral efficiency
of $1.5$~bits/s/Hz is achieved.  For this calculation, we
take the maximum pulse width (containing $99\%$ of the
signal energy) and the maximum bandwidth of the signal set.
By increasing $n$ and $m$, pulse widths get large and the shift of the signal energy from
the symbol period, due to the Gordon-Haus effect, becomes less
significant. Gordon-Haus effect for solitons is as
important as it is for $\sinc$ function transmission
and backpropagation.

We would note that the spectral efficiency reported here was
achieved using only a rather simplistic design approach.  We
believe that a more sophisticated search over the design
space, in particular exploiting the possibility of more
cleverly modulating spectral amplitudes and phase and
choosing eigenvalues in a region not limited to the
$j\omega$ axis, is likely to yield significantly higher
spectral efficiencies.

The recent work of \cite{feder2012} also describes
optical transmission schemes based on $N$-soliton
transmission and the inverse scattering transform, with reports
on some achievable spectral efficiencies.

\subsection{Spectral Efficiencies Achievable by Modulating
the Continuous Spectrum}
\label{sec:contin}

In addition to the discrete spectrum, the continuous
spectrum can, in some cases, also be modulated.

Here we consider the modulation of classical raised-cosine
pulses, with 50\% excess bandwidth.  The continuous spectrum
of an isolated raised-cosine pulse is purely continuous at
low amplitudes, resembling its ordinary Fourier transform.
We modulated the amplitude of an isolated raised-cosine
pulse, propagated the pulse over an optical fiber channel, and
estimated the continuous spectrum of the received signal.
The received spectrum was then compared with the spectrum of
all possible transmitted waveforms at the transmitter using
the log-Euclidean distance
\begin{IEEEeqnarray*}{rCl}
  d(\hat{q}_2(\lambda),\hat{q}_1(\lambda))=\frac{1}{\pi}\int\limits_{-\infty}^{\infty}
\log\left(1+\left|\hat{q}_2(\lambda)-\hat{q}_1(\lambda)\right|^2\right)\der\lambda.
\end{IEEEeqnarray*} 

Fig.~\ref{fig:capacity-contin}(a) shows the estimated achievable rates
in a typical single-channel fiber-optic system, comparing
detection after filtering,
backpropagation, and matched-filtering,
with detection using the nonlinear Fourier transform.
A multi-ring phase-shift keying modulation was used, with rings at 16 distinct amplitude values and 
32 different phase values per ring. The complex
plane was quantized into the Voronoi regions corresponding
to the ring constellation,  to discretize the channel input. Capacity was
computed via the Arimoto-Blahut algorithm.
The NFT
is calculated at either 64 or 1024 uniformly spaced discrete points on
the real axis over a range containing most of the pulse
energy.
The 1024-point NFT is compared with a 1024-point fast Fourier transform (FFT)
implementation of the split-step Fourier method in
the backpropagation scheme. The simulation is performed for a standard single-mode fiber with
dispersion parameter  $D=17~\rm{ps/(nm-km)}$. As can be seen, the NFT and
backpropagation methods achieve approximately the same
rates.
The slight improvement in the NFT method can be attributed to
the stability of the continuous spectral data compared to the
traditional time data (amplitude and phase at the output of
a matched filter).

This simulation was repeated for 5 WDM channels with the
system architecture of Fig.~\ref{fig:fiber-system}. Here,
low-pass filters and ROADMs are placed at the end of each
fiber segment. The spectral efficiency in this case is
obviously lower due to the inter-channel interference. Here
too, NFT and FFT-based backpropagation produce approximately
the same results.

From Fig.~\ref{fig:capacity-contin} it follows that at low \SNRs\
NFT and backpropagation achieve about the same rates. As
the \SNR\ is
increased, the spectral efficiency of backpropagation degrades due
to ISI (under the commonly-assumed isolated symbol detection, in which the memory is 
not accounted for) or inter-channel interference (in a network scenario). However, the 
spectral efficiency of the NFT  scheme may be higher due to its inherent immunity to cross-talk, 
provided that users are multiplexed appropriately in the
nonlinear frequency domain in a multiuser setup; see Section~\ref{sec:nfdm-rates}.

We have not yet
simulated the spectral efficiency at \SNRs\ beyond those shown in
Figs.~\ref{fig:capacity-contin} (a)--(b) due to  high numerical
complexity. Up to the \SNR s $25-30$ dB shown in Figs. \ref{fig:capacity-contin} (a)--(b)
signals only have a continuous spectrum. This made simulations doable. Beyond these \SNR s, discrete mass points
start to appear. This is in fact the SNR level where the nonlinearity starts
to become significant (because, in the focusing regime, the discrete spectrum is the component of the NFT which
is primarily responsible for the nonlinearity). 

In the $N$-soliton transmission simulations of the previous subsections, we began with
a desired spectral constellation at the transmitter. As a result, the initial conditions needed in the Newton algorithm 
at the receiver were known and the detection was computationally feasible. The difficulty there, however, was that
the properties of the resulting signals in the time-domain (bandwidth and time duration) were not properly understood. In contrast, in 
this subsection we started with a signal set in the time domain  \ie, raised-cosine signals. When a signal is
scaled according to a time-domain constellation, discrete mass points appear at unknown places in the complex plane. The difficulty 
here is that the properties of the resulting signals in the spectral domain are not properly understood. As a result, it is
difficult to search for these discrete mass points without a priori information on their locations. The potential 
advantage of NFDM thus remains to be illustrated.

\section{Discussion}
\label{sec:remarks}

In this section, we make a few remarks about the NFT method,
clarify some of its properties and potential practical
issues, and suggest some possible
directions for future investigation.

\subsection{Multiuser Communications Using the NFT}

\begin{figure}[tbhp]
\centering
\includegraphics{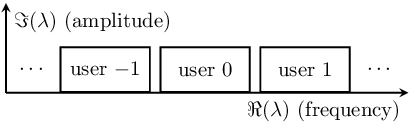}
\caption{ Partitioning $\Complex^+$ for multiuser communication using the NFT.}
\label{fig:multiuser-nft}
\end{figure}

A potential gain is achievable in fiber-optic systems by employing 
methods, such as NFDM, which are less prone to inter-channel interference. 

Recall that in NFDM the real and imaginary axes of the
complex plane correspond, approximately, to the signal
frequency and energy. To use NFDM as a multiuser
communication method, we can partition the complex plane
into disjoint regions, \eg, vertical bins. Each bin can be
assigned to a user and can contain one or more
degrees-of-freedom. To multiplex user signals, traditional
ROADMs must be replaced with nonlinear add-drop multiplexers
(NADMs) which function according to the NFT. In principle,
each NADM would compute the spectrum of its input signals and
filter the signals to be dropped in $\Complex^+$. It would then places the spectrum of the
signals to be added in empty bands and produces the output
signal by taking the inverse NFT. In this way, each user is assigned a
region in the complex plane and, in the absence of noise, does not interfere with
other users; see Fig.~\ref{fig:multiuser-nft} as well as Section~\ref{subsec:noise}. In such a manner, NFDM 
results in an orthogonalization for the deterministic nonlinear Schr\"odinger channel.

\subsection{Noise in the Spectral Coordinates}
\label{subsec:noise}

In a nonlinear interference channel, the interference
can have two components.  The first, termed
``deterministic interference,''  arises from the
(deterministic) interaction of the signals of other users with the
user of interest, and in general is present even in the
absence of noise.  The second, termed ``stochastic interference,''
arises from the (stochastic) coupling of noise with the
signals of other users, interfering with the channel of
interest.  Thus, noise can affect the channel of interest
directly (in-band noise) and indirectly (by
introducing interference). Typically deterministic interference 
is stronger than stochastic interference.

The NLS equation with additive noise has no known
integrability structure, in the sense of possessing a set of
non-interacting degrees-of-freedom. As a result, while the NFT method 
does not suffer from (strong) deterministic interference,
a (weak) stochastic interference is expected to be present.
In other words, even when users are multiplexed so that they
do not interact in the absence of noise, the addition of
noise can introduce stochastic interference.
In comparison, conventional WDM with backpropagation is
subject to both deterministic and stochastic interference.   

Finally, note that an additive (in-band) white noise in the time domain has
coordinates in the spectral domain which may not be independent.
Such correlations should be accounted for when designing
signal detectors.

\subsection{NFDM versus OFDM}

NFDM and OFDM are essentially similar in the absence of noise.
However, as noted above,
in
cases where the presence of noise breaks the integrability
structure, then, unlike OFDM, the signal degrees-of-freedom
are not independent in NFDM.

Note also that, while the ordinary Fourier transform of a
signal is a function of a real variable, the NFT of a signal
is generally defined on the whole complex plane, \ie ,
nonlinear frequencies are complex-valued. Since complex
frequencies in the upper half complex plane are
isolated points, this component in the support of the NFT
can be modulated too. OFDM is conceptually in analogy with
the continuous spectrum modulation, where only the spectral
amplitudes are modulated.  The discrete spectrum differs
further from the continuous spectrum in that it is the
frequencies in $\Complex^+$ themselves that contribute to
the signal energy, and not their spectral amplitudes.

The analogy between NFDM and OFDM can be better understood in fibers
with negative dispersion parameter. In this case, $q^*$ is replaced 
with $-q^*$ in \eqref{eq:L-op-part3} and \eqref{eq:dv-dt-part3} and the (noise-free) channel is
still integrable. Here, similar to OFDM, the nonlinear frequencies are 
real and information is encoded only in spectral amplitudes. This case is
appealing analytically and numerically, since the underlying $L$
operator is Hermitian.

\subsection{Advantages and Disadvantages of NFDM}
\subsubsection{Advantages}
Some of the advantages of using NFDM were outlined in Section~\ref{sec:intro}.
In short, deterministic 
distortions, \ie, distortions that arise even in
a noise-free system (SPM, XPM, FWM, ISI and interference) are zero for all 
users of a multiuser network.
\subsubsection{Disadvantages}

\paragraph{Aplicable only to Perfectly Integrable Models}
NFDM critically relies on the integrability of the channel.
Loss, higher-order dispersion, and other perturbations caused by filters and
  communication equipment not taken into account in this study contribute to
deviations from integrability. There are, however, several reasons to
believe that the overall channel from the transmitter to the receiver
can still be close to an integrable channel:
\begin{itemize}
\item Using Raman amplification, the effects of loss are
  minimal (and indeed are traded with noise perturbation). 
\item The NFT is applicable as long as the system can support soliton
transmission.
  Solitons have been implemented in practice in the presence of
  communication equipment (filters, multiplexers,
  analog-to-digital (A/D) converters, etc). This is
  an indicator that the overall channel is still nearly integrable.   
\item Mathematically one has stability results for solitons
\cite{tao2009soliton}. A soliton
  passing through a filter might be slightly distorted, but it
  re-organizes its shape so as to revert back to its
original shape (or to form a soliton with a nearby discrete
spectrum in the complex plane). 
\item Considering that the performance of the WDM method degrades asymptotically
with \SNR , it may be worthwhile to identify sources of perturbations to integrability in NFDM and minimize them. This
way, rather than engineering the channel to be linear,
the near-integrable channel is engineered to be integrable, so as to support its natural nonlinear modes, 
which characteristically tend to form in the medium.
\end{itemize}

\paragraph{High Complexity}
The nonlinear Fourier coefficients at the receiver
  are calculated using $\bigo(n)$ operations per nonlinear frequency,
  where $n$ is the number of signal samples in time. In comparison with
  with the $\bigo(n\log n)$ complexity of the FFT, the complexity of an
  $n$-point NFT is, at present,
  $\bigo(n^2)$ (assuming a fixed number of Newton steps are needed in the
case of discrete frequencies). The complexity of the transmitter can be even
  higher. As a result, the NFT is currently computationally difficult to
implement or sometimes even to simulate.  It is therefore of interest to
develop faster algorithms.

\paragraph{Hardware Limitations}
Optical or electrical signal processing of $N$-solitons and their
  required hardware (\eg , A/Ds) may not be as simple as those in linear
  systems. The NFT decoder typically requires
  signal samples at increments smaller than the Nyquist period. An
  interpolation step may be needed to find all the necessary data.

\subsection{Achievable Rates Using NFDM}
\label{sec:nfdm-rates}
As mentioned earlier, a stochastic interference can be present in NFDM.
As a result, although the achievable rate of NFDM in a multiuser channel can be 
higher than that of WDM with backpropagation (at least in the limit that noise 
goes to zero), it too may ultimately 
peak at some finite \SNR. We have not yet
simulated the capacity at high \SNRs\ to see when this
may potentially happen. However, note that, regardless of the
method of transmission, the noisy channel
will fundamentally be subject to interference due to the lack of
integrability\footnote{Since, for instance, the NLS equation with a generic additive
  potential does not have conserved quantities.}.

Note that the decline of the achievable rates in WDM simulations in
prior work is mostly due to the deterministic distortions. Our capacity simulations, 
as well as those in \cite[Fig. 36]{essiambre2010clo}, show that in the absence of noise the rates 
of the WDM method still vanish (or saturate, depending on assumptions
on interfering users) at high powers in the network scenario. This is not the 
case for multiuser NFDM whose
achievable rate is unbounded when noise is absent. We thus expect
improvements in the information rates using NFDM, due to the
immunity to the deterministic distortions. 

\subsection{Eigenvalue Communication of Hasegawa and Nyu}
\label{sec:eig-com}
As noted in introduction, as well as in [Part~I],  the work of Hasegawa and Nyu on ``eigenvalue
communication'' \cite{hasegawa1993ec} is related to the
NFT-based approach taken in this paper.  Hasegawa and Nyu make the 
observation that the time-domain data is distorted while
the eigenvalues remain constant and can be used for communication. 
The authors then considered single-user channels, encoded information
in conserved quantities, and used the inverse scattering transform (IST)
as a means to decode these conserved quantities. The idea is
illustrated for pulses of the form $A\sech(t)$.

The use of conserved quantities in a communication channel is desirable since it
facilitates deterministic signal processing at the receiver and may
simplify communication design. However, fundamentally it does not
offer any capacity 
improvement if one uses an equivalent set of
non-conserved quantities (\eg , amplitude and phase). In fact, it is clear that the 
\emph{entire signal itself} is a conserved quantity under backpropagation and can 
be perfectly recovered. Thus the approach of \cite{hasegawa1993ec} does not achieve anything more than
what BP does. It is therefore not
necessary to aim at extracting and modulating quantities
conserved by a channel. Furthermore, the application of conserved quantities in a multiuser
setup is in and of itself not useful, since conservation does not imply separability which underlies
the NFDM approach.

In contrast, our primary motivation for introducing
NFT-based methods stems from recent capacity studies
\cite{essiambre2010clo} and our observation, made
in Section~\ref{sec:cap-limit}, that a major reason that the capacity rolls off in prior work, 
after abstracting away non-essential aspects, is predominantly 
because these methods, in essence, modulate linear-algebraic modes. When used in a nonlinear 
channel, this leads to a significant inter-channel interference and ISI.  
We noted that the NLS equation, however, supports nonlinear modes which have
a crucial \emph{independence property}, that can be used to build a interference-free multiuser 
system. The mathematical framework necessary to reveal signal degrees-of-freedom
is the IST. As a result, in [Part~I] we began with IST, shifted the focus from the scattering theory, 
considered IST as a Fourier transform,  filtered its literature accordingly and presented a signal-and-systems perspective.
This then paved the way for [Part II] and this paper whose overall goal was to construct NFDM,
which can be viewed as a generalization of OFDM to nonlinear systems. The existence of an OFDM-like
scheme for a nonlinear system is rather surprising. In [Part~II] and this paper, we developed details of  
NFDM transmitter and receiver. Note that NFDM may not have eigenvalues as in \cite{hasegawa1993ec}. Indeed 
the analogy between NFDM and OFDM can be best understood in the defocusing regime where,
similarly to the ordinary Fourier transform, one has a transformation $\hat q(\lambda)$. This signal 
transformation underlies NFDM.

\subsection{Linear Multiplexing Methods other than WDM}
\label{sec:tdm-ofdm-mmf}
In this paper, we mostly focused on WDM as an example a linear multiplexing method. 
There are however other methods as well which, in essence, modulate 
linear-algebraic modes for transmission and behave similarly. Time-division multiplexing (TDM), 
OFDM and multi-mode communication are other examples.

Although mathematically these methods are not exact orthogonalizations, within a certain range of system
 and user parameters they can potentially be approximate orthogonalizations. For
 instance, at low powers where the nonlinearity is weak, each of $n$ WDM users achieves $1/n$ of the degrees-of-freedom.
Among these schemes, TDM is different in that, the dimension in which the multiplexing occurs, \ie, time, can 
be practically unlimited. TDM can be capacity-achieving if a
 sufficient guard time, depending on the users' powers and transmission distance, is
 introduced between users' signals. However, a system designed for one transmission
 distance may fail to operate at a larger distance: As the signal propagates further in
 distance, at some point orthogonalization is lost. Also, in principle, as the signal power
tends to infinity, the transmission time of the user of interest tends to infinity as well ---
in effect, TDM turns the multiuser channel into a single-user one in order to address the
interaction problem associated with the linear multiplexing. Ignoring practical issues
in a network scenario, a TDM system designed for the worst-case system and
user parameters can be nearly capacity-achieving in that regime.
In fact, this is also true for WDM: if one has infinite bandwidth, users' signals can be widely separated such
that in the range of system and user parameters their interaction is negligible.

The distinguishing feature of NFDM is that, deterministically, it is an exact orthogonalization for any
transmission distance, signal power, dispersion or nonlinearity parameter. Users' signals may overlap in
time and frequency, but they are separated in the nonlinear Fourier domain.

\subsection{The Discrete Nonlinear Fourier Transform}
To implement NFDM in practice, the discrete nonlinear Fourier
transform, in which time domain signal is discrete and periodic, should be implemented.
The development of the discrete nonlinear Fourier transform exists in 
the mathematics literature \cite{ma1981pcs,tracy1984,tracy1988nsm, osborne2009now} --- although it is not as fully developed 
as the continuous one. There are also important differences in the way that the spectrum is defined.

\section{Conclusions}
\label{sec:conclusions}
Motivated by recent studies showing that the achievable
rates of current methods in optical fiber networks vanish at high
launch power due to the impact of nonlinearity, in [Part~I],
[Part~II], and this paper we have revisited
information transmission in such nonlinear systems.
In these papers, we suggested
using the nonlinear Fourier
transform to transmit information over integrable communication
channels such as the optical fiber channel, which is
governed by the nonlinear Schr\"odinger equation.
In this transmission scheme, information is encoded in the
nonlinear Fourier transform of the signal, consisting of two
components: a discrete and a continuous spectral function.
With this new method, deterministic
distortions arising from the dispersion and nonlinearity, such as
inter-symbol and inter-channel interference are zero 
for a single-user channel or all users of a multiuser network.

We took the first steps towards the design of a
communication system implementing the nonlinear Fourier
transform. We proposed a Darboux-transform-based algorithm
for modulating the discrete spectrum at the transmitter,
and we provided a first-order perturbation analysis of the
influence of noise on the received spectrum.
Furthermore,
we provided examples illustrating how the NFT can
be used for data transmission. Although these small examples
clearly demonstrate improvements over their benchmark
systems, more sophisticated large-scale simulations are
required to demonstrate the potential to achieve high
spectral efficiencies.

Because nonlinearity is a key feature of fiber-optic
networks,
the development of nonlinearity-compatible
transmission schemes, like those based on the nonlinear Fourier
transform, is likely to continue to be a fruitful research
direction.

\section*{Acknowledgment}
The authors would like to thank Ren\'e-Jean~Essiambre, Aris L.
Moustakas and Andrew C. Singer for helpful comments at
various stages of this work, and especially Gerhard Kramer
for his helpful comments
and much discussion on the fiber-optic transmission problem.

\appendices
\section{Solution of Hirota Equations}
\label{app:hirota}

Because (\ref{eq:FG1}) and (\ref{eq:FG2}) are homogeneous equations in
the order of derivatives that occur in each term,
exponential functions are candidate solutions.  Let us expand
$F$ and $G$ as
\begin{IEEEeqnarray*}{rCl}
 F(t,z)&=&f_0(t,z)+\epsilon f_1(t,z)+\epsilon^2 f_2(t,z)+\cdots,\\
G(t,z) &=&  g_0(t,z)+\epsilon g_1(t,z)+\epsilon^2 g_2(t,z)+\cdots,
\end{IEEEeqnarray*}
for some small parameter $\epsilon$.  The bilinear terms in
\eqref{eq:FG1}--\eqref{eq:FG2} are 
\begin{IEEEeqnarray*}{rCl}
FG&=&\sum\limits_{n=0}^{\infty}\epsilon^n\left(\sum\limits_{k=0}^nf_{k}g_{n-k}\right), \: FF=\sum\limits_{n=0}^{\infty}\epsilon^n\left(\sum\limits_{k=0}^nf_{k}f_{n-k}\right),\\
|G|^2&=&\sum\limits_{n=0}^{\infty}\epsilon^n\left(\sum\limits_{k=0}^ng_{k}g_{n-k}^*\right).
\end{IEEEeqnarray*}

Substituting these expressions into
\eqref{eq:FG1}--\eqref{eq:FG2} and equating like powers of
$\epsilon$ gives rise to a recursive procedure
to obtain $\{f_{n+1}, g_{n+1}\}$ from  $\{f_n, g_n\}$. 
To begin finding unknowns recursively, we can set initially $f_0=1$. 
The zero-order term $\epsilon^0$ then gives $g_0=0$. Starting with
$\{f_0=1, g_0=0\}$, the recursive equations are:
\begin{IEEEeqnarray*}{rCl}
\begin{cases}
H 1.g_n= -\sum\limits_{k=1}^{n-1}H(f_kg_{n-k}),  & \\
\partial_{tt}f_n=-\frac{1}{2}\sum\limits_{k=1}^{n-1}D_t^2f_kf_{n-k} +\gamma \sum\limits_{k=1}^{n-1}g_kg_{n-k}^* .&
\end{cases}
\end{IEEEeqnarray*}
where $H=jD_z+D_{t}^2$. It can be seen that $f_{2n+1}=g_{2n}=0$. At each iteration $n$ one of
$f_n$ or $g_n$ is non-zero, which is used in the next iteration.
For a general nonlinear system, iterations continue
indefinitely. However, for integrable equations, as shown below for
the NLS equation,  this series truncates and exact solutions of various finite
order are obtained. 

Before proceeding, it is useful to know properties of the Hirota
operators. Let $X_i=\zeta_i t-k_iz+\phi_i$, with
dispersion relation $k_i=j\zeta_i^2$. Then:
 
\begin{enumerate}
\item
$D_t^2e^{X_i}e^{X_j}=D_t^2e^{X_j}e^{X_i}=(\zeta_j-\zeta_i)^2e^{X_i+X_j}$;

\item
$D_ze^{X_i}e^{X_j}=-D_ze^{X_j}e^{X_i}=(k_j-k_i)e^{X_i+X_j}$; 

\item
$He^{X_i}e^{X_j}=\left(j(k_j-k_i)+(\zeta_j-\zeta_i)^2
\right)e^{X_i+X_j}$;

\item  $H(1.e^X)=(jk+\zeta^2)e^X$, $X=\zeta t-k z+\phi$;

\item  $H(e^{X_i}e^{X_j})=\alpha_{ij} H (1.e^{X_i+X_j})$,
where
$\alpha_{ij}=\left( j(k_j-k_i)+(\zeta_j-\zeta_i)^2\right)/\left(j(k_j+k_i)+(\zeta_j+\zeta_i)^2 \right)$;

\item
 $H(e^{X_i^*}e^{X_i^*+X_j})=H(e^{X_i+X_j^*}e^{X_i})=0$; 
\item  $H(1.g)=He^{X_i}e^{X_j}$ has a solution $g=\alpha_{ij} e^{X_i+X_j}$.
\end{enumerate}

The first iteration $n=1$ is 
\begin{IEEEeqnarray*}{rCl}
\begin{cases}
H 1.g_1=0,\\
(f_1)_{tt}=0,
\end{cases}
\end{IEEEeqnarray*}
which gives
\begin{IEEEeqnarray*}{rCl}
g_1=\sum\limits_{i=1}^N e^{X_i},\quad f_1=0,
\end{IEEEeqnarray*}
where $N\in\mathbb{N}$ is the order of the multi-soliton solution.

We first consider $N=1$ and then show how an additional mass point can be added to
obtain the $N=2$ solution.

\subsection{Case N=1}
We have $f_0=1$, $g_0=f_1=0$ and $g_1=e^{X_1}$. For $n=2$ 
\begin{IEEEeqnarray*}{rCl}
\begin{cases}
  H(1.g_2)=-H (f_1g_1)=0,  \\
  (f_2)_{tt}=-\frac{1}{2}D_t^2 f_1f_1+ \gamma g_1g_1^*= \gamma e^{X_1+X_1^*}.
\end{cases}
\end{IEEEeqnarray*}
At this point, since one variable $X_1$ already exists and $N=1$, we aim at truncating
the process by choosing zero solutions and avoiding the inclusion of terms $\exp(X_i)$, $i\geq 2$.
We thus consider the solution
\begin{IEEEeqnarray*}{rCl}
  g_2 = 0 ,  \quad f_2 = e^{X_1+X_1^*+R_{13}},
\end{IEEEeqnarray*}
where $R_{13}=-2\log(\zeta_1+\zeta_1^*)+\log\gamma$. It can be verified that the above solution then gives all
other terms $f_k=g_k=0$, $k\geq 3$, and the iteration truncates. The
1-soliton solution is obtained as
\begin{IEEEeqnarray*}{rCl}
q=\frac{G}{F}&=&\frac{e^{X_1}}{1+e^{X_1+X_1^*+R_{13}}}
\\
&=&\frac{1}{2}e^{-\frac{1}{2}R_{13}} e^{j\Im X_1}\sech\left(\Re
  X_1+\frac{1}{2}R_{13}\right).
\end{IEEEeqnarray*}
By setting $\zeta_1=-2j\lambda_1=\omega-j\alpha$ where
$\lambda_1=(\alpha+j\omega)/2$ is the eigenvalue, $k_1=j\zeta_1^2=2\alpha\omega-j(\alpha^2-\omega^2)$, $\phi_1=\log
\frac{2\omega^2}{\tilde{q}}$,
we get
\begin{IEEEeqnarray*}{rCl}
q=\omega e^{-j\alpha t +j(\alpha^2-\omega^2)z-j\angle \tilde{q}}
\sech\left(\omega(t-2\alpha z-t_0)\right),
\end{IEEEeqnarray*}
where $t_0=\frac{1}{\omega}\log\frac{\gamma|\tilde q|}{\omega}$ and $\angle$ denotes phase.

\subsection{Case N=2}
For $n=1$ we get $f_0=1$, $g_0=f_1=0$ and
$g_1=e^{X_1}+e^{X_2}$. The next iteration $n=2$ is 
\begin{IEEEeqnarray*}{rCl}
  \begin{cases}
  H (1. g_2) = -H f_1g_1=0,  \\
  (f_2)_{tt}=-\frac{1}{2} D_t^2 f_1f_1 +  \gamma g_1g_1^* \\
\hspace{0.92cm} = \gamma \left(e^{X_1+X_1^*}+e^{X_1+X_2^*}+e^{X_2+X_1^*}+e^{X_2+X_2^*}\right).
\end{cases}
\end{IEEEeqnarray*}
We choose the solution   $g_2=0$ and
\begin{IEEEeqnarray*}{rCl}
  f_2&=&e^{X_1+X_1^*+R_{13}}+e^{X_1+X_2^*+R_{14}}+e^{X_2+X_1^*+R_{23}}\\
&&+\:e^{X_2+X_2^*+R_{24}},
\end{IEEEeqnarray*}
where 
\begin{equation*}
R_{ij}=\begin{cases}
2\log(\zeta_i-\zeta_j)-\log\gamma,& i, j \leq N \mbox{ or } i,j\geq N+1,\\
-2\log(\zeta_i+\zeta_j)+\log\gamma, & i\leq N\mbox{ and } j\geq N+1,
\end{cases}
\end{equation*}
and $\zeta_{i+N=}\zeta_i^*$.

The $n=3$ iteration is
\begin{IEEEeqnarray*}{rCl}
\begin{cases}
 H1.g_3= -H(f_1g_2+ f_2g_1)=-Hf_2(e^{X_1}+e^{X_2}),
 \\
  \partial_{tt}f_3=0.
\end{cases}
\end{IEEEeqnarray*}
The second equation gives $f_3=0$. The first one, using property 6), simplifies to
\begin{IEEEeqnarray*}{rCl}
H (1.g_3)&=&-H(e^{X_1^*+X_2+R_{23}}.e^{X_1})-H(e^{X_2+X_2^*+R_{24}}.e^{X_1})\\
&&-
H(e^{X_1^*+X_1+R_{13}}.e^{X_2})-
H(e^{X_1+X_2^*+R_{14}}.e^{X_2}).
\end{IEEEeqnarray*}
Using property 7), a solution is
\begin{IEEEeqnarray}{rCl}
  g_3&=&  -\Bigl(\alpha_{23,1}e^{X_1+X_1^*+X_2+R_{23}}+
\alpha_{24,1}e^{X_1+X_2+X_2^*+R_{24}}\nonumber\\
&&+
\alpha_{13,2}e^{X_1+X_1^*+X_2+R_{13}}
+\alpha_{14,2}e^{X_1+X_2^*+X_2+R_{14}}\Bigr)
\nonumber\\
&=&
-\left(\alpha_{13,2}e^{R_{13}}+\alpha_{23,1}e^{R_{23}}\right)
e^{X_1+X_1^*+X_2}\nonumber\\
&&-
\left(\alpha_{14,2}e^{R_{14}}+\alpha_{24,1}e^{R_{24}}\right)
e^{X_1+X_2+X_2^*}.
\label{eq:N=2-g3-intermed}
\end{IEEEeqnarray}
Here
\begin{IEEEeqnarray*}{rCl}
\alpha_{(i+N)j,i}&=&
\frac{j(k_i-(k_i^*+k_j)))+(\zeta_i-(\zeta_i^*+\zeta_j)^2 }{j(k_i+(k_i^*+k_j)))+(\zeta_i+(\zeta_i^*+\zeta_j)^2}\\
&=&
\frac{2\zeta_j^2 -2|\zeta_i|^2-2\zeta_i\zeta_j+2\zeta_i^*\zeta_j}{ 2\zeta_i^{*2}+2|\zeta_i|^2+2\zeta_i\zeta_j+2\zeta_i^*\zeta_j}
\\
&=&
\frac{\zeta_j-\zeta_i}{\zeta_i+\zeta_i^*},
\end{IEEEeqnarray*}
and similarly,
\begin{IEEEeqnarray*}{rCl}
\alpha_{(i+N)i,j}&=&
\frac{j(k_j-(k_i^*+k_i)))+(\zeta_j-(\zeta_i^*+\zeta_i))^2 }{j(k_j+(k_i^*+k_i)))+(\zeta_j+(\zeta_i^*+\zeta_i))^2}\\
&=&
\frac{\zeta_i-\zeta_j}{\zeta_i^*+\zeta_j}.
\end{IEEEeqnarray*}
The first coefficient $a=\alpha_{13,2}e^{R_{13}}+\alpha_{23,1}e^{R_{23}}$ in \eqref{eq:N=2-g3-intermed} is 
\begin{IEEEeqnarray*}{rCl}
a
&=&
\frac{\zeta_1-\zeta_2}{\zeta_1^*+\zeta_2}e^{R_{13}}+
\frac{\zeta_2-\zeta_1}{\zeta_1+\zeta_1^*}e^{R_{23}}
\\
&=&
\frac{\zeta_1-\zeta_2}{\zeta_1^*+\zeta_2}\times
\frac{\gamma}{(\zeta_1^*+\zeta_1)^2}
+
\frac{\zeta_2-\zeta_1}{\zeta_1+\zeta_1^*}\times
\frac{\gamma}{(\zeta_2+\zeta_1^*)^2}
\\
&=&-
\frac{\gamma(\zeta_1-\zeta_2)^2}{(\zeta_1^*+\zeta_1)^2(\zeta_1^*+\zeta_2)^2}
\\
&=&
-e^{R_{12}+R_{13}+R_{23}}.
\end{IEEEeqnarray*}
In the same manner, the second coefficient $b=\alpha_{14,2}e^{R_{14}}+\alpha_{24,1}e^{R_{24}}$ in \eqref{eq:N=2-g3-intermed} 
is obtained as
\begin{IEEEeqnarray*}{rCl}
b&=&
\frac{-\gamma(\zeta_1-\zeta_2)^2}{(\zeta_1+\zeta_2^*)^2(\zeta_2+\zeta_2^*)^2}=
-e^{R_{12}+R_{14}+R_{24}}.
\end{IEEEeqnarray*}

It follows that 
\begin{IEEEeqnarray*}{rCl}
  g_3=e^{X_1+X_1^*+X_2+R_{12}+R_{13}+R_{23}}+
e^{X_1+X_2+X_2^*+R_{12}+R_{14}+R_{24}}.
\end{IEEEeqnarray*}

The $n=4$ iteration is:
\begin{IEEEeqnarray*}{rCl}
\begin{cases}
  H1.g_4=0,\\
(f_4)_{tt}=-\frac{1}{2}D_t^2(f_2f_2)+\gamma(g_3g_1^*+g_1g_3^*).
\end{cases}
\end{IEEEeqnarray*}
The first equation gives $g_4=0$. For the second one, using property
1), we have
\begin{IEEEeqnarray}{rCl}
\frac{1}{2}D_t^2f_2.f_2&=&\frac{1}{2}D_t^2\Bigl(e^{X_1+X_1^*+R_{13}}+e^{X_1+X_2^*+R_{14}}
\nonumber\\
&&+\: e^{X_2+X_1^*+R_{23}}+e^{X_2+X_2^*+R_{24}}\Bigr)^2
\nonumber\\
&=&
(\zeta_2^*-\zeta_1^*)^2e^{2X_1+X_1^*+X_2^*+R_{13}+R_{14}}\nonumber\\
&&+\:
(\zeta_2-\zeta_1)^2e^{X_1+2X_1^*+X_2+R_{13}+R_{23}}\nonumber\\
&&+\:
(\zeta_2+\zeta_2^*-\zeta_1-\zeta_1^*)^2e^{X_1+X_1^*+X_2+X_2^*+R_{13}+R_{24}}\nonumber\\
&&+\:
(\zeta_1^*+\zeta_2-\zeta_1-\zeta_2^*)^2e^{X_1+X_1^*+X_2+X_2^*+R_{14}+R_{23}}
\nonumber\\
&&+\:
(\zeta_2-\zeta_1)^2e^{X_1+X_2+2X_2^*+R_{14}+R_{24}}\nonumber\\
&&+\:
(\zeta_2^*-\zeta_1^*)^2e^{X_1^*+2X_2+X_2^*+R_{23}+R_{24}}
\nonumber\\
&=&
(\zeta_2+\zeta_2^*-\zeta_1-\zeta_1^*)^2e^{X_1+X_1^*+X_2+X_2^*+R_{13}+R_{24}}\nonumber\\
&&+\:
(\zeta_1^*+\zeta_2-\zeta_1-\zeta_2^*)^2e^{X_1+X_1^*+X_2+X_2^*+R_{14}+R_{23}}
\nonumber\\
&&+\:\gamma
\Bigl\{
e^{2X_1+X_1^*+X_2^*+R_{13}+R_{14}+R_{34}}\nonumber\\
&&+\:
e^{X_1+2X_1^*+X_2+R_{13}+R_{23}+R_{12}} \nonumber\\
&&+\:
e^{X_1+X_2+2X_2^*+R_{14}+R_{24}+R_{12}}\nonumber\\
&&+\:
e^{X_1^*+2X_2+X_2^*+R_{23}+R_{24}+R_{34}}\Bigr\}.
\label{eq:f2-f2}
\end{IEEEeqnarray}
Also
\begin{IEEEeqnarray}{rCl}
\gamma(g_3g_1^*+\text{c.c.})&=&\gamma\Bigl[
(e^{R_{13}+R_{23}}+e^{R_{14}+R_{24}})e^{R_{12}}+\text{c.c.}\Bigr]
\nonumber \\
&& \times\: e^{X_1+X_1^*+X_2+X_2^*}
\nonumber\\
&&+\:\gamma
\Bigl\{
e^{X_1+2X_1^*+X_2+R_{12} + R_{13}+R_{23} } \nonumber\\
&&+\:
e^{X_1+X_2+2X_2^*+R_{12}+R_{14}+R_{24}}
\nonumber\\
&&
+\:
e^{2X_1+X_1^*+X_2^*+R_{12}^* + R_{13}^*+R_{23}^* } 
\nonumber\\
&&+\:
e^{X_1^*+2X_2+X_2^*+R_{12}^*+R_{14}^*+R_{24}^*}\Bigr\}.
\label{eq:g3-g1}
\end{IEEEeqnarray}
The terms inside braces in \eqref{eq:f2-f2} and \eqref{eq:g3-g1},
involving $2X_1$, $2X_2$, $2X_1^*$ and $2X_2^*$, cancel out using
identities 
$R_{12}^*=R_{34}$, $R_{13}^*=R_{13}$, $R_{14}^*=R_{23}$, $R_{24}^*=R_{24}$.
The terms containing $\exp(X_1+X_1^*+X_2+X_2^*)$, after some algebra, simplify and we obtain
\[
f_4=e^{X_1+X_1^*+X_2+X_2^*+R_{12}+R_{13}+R_{14}+R_{23}+R_{24}+R_{34}}.
\]

One can check that $f_k=g_k=0$, $k\geq 5$. It follows that
\begin{IEEEeqnarray*}{rCl}
G&=&e^{X_1}+e^{X_2}+e^{X_1+X_1^*+X_2+R_{12}+R_{13}+R_{23}}\\
&&+\:
e^{X_1+X_2+X_2^*+R_{12}+R_{14}+R_{24}},\\
F
&=&1+e^{X_1+X_1^*+R_{13}}+e^{X_1+X_2^*+R_{14}}+e^{X_2+X_1^*+R_{23}}\\
&&+\: e^{X_2+X_2^*+R_{24}}\\
&&+\:e^{X_1+X_1^*+X_2+X_2^*+R_{12}+R_{13}+R_{14}+R_{23}+R_{24}+R_{34}}.
\end{IEEEeqnarray*}

\subsection{General Formula}
The calculations in the previous subsections illustrate how one can add an additional term
$\exp(X_N)$ to $g_1$ and update $F$ and $G$. Using similar calculations,
$N$-soliton solutions are obtained inductively. Furthermore, at this point 
the structure of $F$ and $G$, noted earlier, becomes clear.
 
In the Hirota method, eigenvalues
$\lambda_i=(\alpha_i+j\omega_i)/2$ are related to $\zeta_i$ by $\zeta_i=-2j\lambda_i=\omega_i-j\alpha_i$. For
$N=1$, the spectral amplitude is determined via $\phi_1=\log(2\omega_1^2/\tilde{q}_1)$. While
eigenvalues in the Riemann-Hilbert, Hirota and Darboux methods are the same, other spectral parameters
are different. 

\section{Proof of the Darboux Theorem}
\label{app:darboux}

See \cite{matveev1991darboux} for the proof of a more general theorem. Here we give a simple proof for 
Theorem~\ref{thm:darboux}.

Let $\phi(t,\lambda;q)$ be a known eigenvector associated with $\lambda$ and
$q$, \ie, satisfying $\phi_t=P(\lambda, q)\phi$. Its adjoint
$\tilde{\phi}(t,\lambda;q)=[\phi_2^*,-\phi_1^*]$ satisfies
$\tilde{\phi}_t=P(\lambda^*, q)\tilde{\phi}$. Denote
this known solution as $S=[\phi , \tilde{\phi}]$,
$\Gamma=\diag(\lambda,\lambda^*)$, and $\Sigma=S\Gamma S^{-1}$.

We can verify that $S_t=JS\Gamma+QS$, where
$J=\diag(-j,j)$ and $Q=\offdiag(q,-q^*)$. In addition we have 
$\Sigma_t=[J\Sigma+Q,\Sigma]$.

Given that $\phi(t,\lambda; q)$ is known, the Darboux transformation
maps $\{v(t,\mu;q),\tilde{v}(t,\mu;q)\}$ to 
$\{u(t,\mu;\tilde{q}),\tilde{u}(t,\mu;\tilde{q})\}$ according to 
\[
U=V\Lambda-\Sigma V,
\]
where $V=[v,\tilde{v}]$, $U=[u, \tilde{u}]$,
$\Lambda=\diag(\mu,\mu^*)$.  

We have $V_t=JV\Lambda+QV$ and
\begin{IEEEeqnarray*}{rCl}
 U_t&=&V_t\Lambda-(\Sigma_t V+\Sigma V_t)\\
&=& (JV\Lambda+QV)\Lambda-\left([J\Sigma+Q, \Sigma]V+\Sigma
  (JV\Lambda+QV)\right)\\
&=&
(JV\Lambda+Q V)\Lambda-\Sigma J
V\Lambda-\left([J\Sigma+Q,\Sigma]+\Sigma Q \right)V
\\
&=&J(V\Lambda-\Sigma V)\Lambda+J\Sigma V\Lambda-\Sigma J V\Lambda \\
&&+\:QV\Lambda-\left([J\Sigma+Q,
     \Sigma]+\Sigma Q\right)V
\\
&=&
JU\Lambda+[J,\Sigma]V\Lambda-\left([J\Sigma+Q,
     \Sigma]+\Sigma Q\right)V+QV\Lambda
\\
&=&
JU\Lambda+[J,\Sigma]V\Lambda-\left(J\Sigma^2+Q\Sigma-\Sigma J\Sigma\right)V
+QV\Lambda
\\
&=&
JU\Lambda+[J,\Sigma]V\Lambda-[J,\Sigma]\Sigma V-Q\Sigma V
+QV\Lambda
\\
&=&
JU\Lambda+[J,\Sigma](V\Lambda-\Sigma V)+Q(V\Lambda-\Sigma V)
\\
&=&JU\Lambda+\left(Q+[J,\Sigma]\right)U
\\
&=&
JU\Lambda+\tilde{Q}U,
\end{IEEEeqnarray*}
where $ \tilde{Q}=Q+[J,\Sigma]$.

It follows that $u$ and $\tilde u$ satisfy the $P$-equations
$u_t=P(\mu, \tilde q)u$ and $\tilde u_t=P(\mu^*, \tilde q)\tilde u$. In the same manner we can show 
that $u$ and $\tilde{u}$ satisfy the
$M$-equations $u_z=M(\mu, \tilde{q})u$ and $\tilde{u}_z=M(\mu^*,\tilde{q})\tilde{u}$.

\end{document}